\documentclass[preprint2]{aastex}

\slugcomment{Accepted for publication in the AJ}

\shorttitle{Variability in PPNs. V. Three O-Rich}

\shortauthors{Hrivnak et al.}

\begin{document}


\title{VARIABILITY IN PROTO-PLANETARY NEBULAE: V. VELOCITY AND LIGHT CURVE ANALYSES OF IRAS 17436+5003, 18095+2704, AND 19475+3119}

\author{Bruce J. Hrivnak\altaffilmark{1}, Griet Van de Steene\altaffilmark{2}, Hans Van Winckel\altaffilmark{3}, Wenxian Lu\altaffilmark{1}, and Julius Sperauskas\altaffilmark{4} }

\altaffiltext{1}{Department of Physics and Astronomy, Valparaiso University, 
Valparaiso, IN 46383; bruce.hrivnak@valpo.edu, wen.lu@valpo.edu (retired)}
\altaffiltext{2}{Royal Observatory of Belgium, Astronomy \& Astrophysics, Ringlaan 3, Brussels, Belgium; g.vandesteene@oma.be}
\altaffiltext{3}{Instituut voor Sterrenkunde, K.U. Leuven University, Celestijnenlaan 200 D, B-3001
Leuven, Belgium; Hans.VanWinckel@ster.kuleuven.be}
\altaffiltext{4}{Astronomical Observatory Vilnius University, Saul\.{e}tekio al. 3, 10257 Vilnius Lithuania; julius.sperauskas@ff.vu.lt}

\begin{abstract}
We have obtained contemporaneous light, color, and radial velocity data for three proto-planetary nebulae (PPNe) over the years 2007 to 2015.  The light and velocity curves of each show similar periods of pulsation, with photometric periods of 42 and 50 days for IRAS 17436+5003, 102 days for IRAS 18095+2704, and 35 days for IRAS 19475+3119.  The light and velocity curves are complex with multiple periods and small, variable amplitudes.  Nevertheless, at least over limited time intervals, we were able to identify dominant periods in the light, color, and velocity curves and compare the phasing of each.  The color curves appear to peak with or  slightly after the light curves while the radial velocity curves peak about a quarter of a cycle before the light curves.  Similar results were found previously for two other PPNe, although for them the light and color appeared to be in phase. Thus it appears that PPNe are brightest when smallest and hottest.   These phase results differ from those found for classical Cepheid variables, where the light and velocity differ by half a cycle, and are hottest at about average size and expanding.  However, they do appear to have similar phasing to the larger amplitude pulsations seen in RV Tauri variables.  Presently, few pulsation models exist for PPNe, and these do not fit the observations well, especially the longer periods observed.  Model fits to these new light and velocity curves would allow masses to be determined for these post-AGB objects, and thereby provide important constraints to post-AGB stellar evolution models of low and intermediate-mass stars.

\end{abstract}


\section{INTRODUCTION}
\label{intro}

Proto-planetary nebulae (PPNe) are objects in the short-lived (few thousand years) evolutionary stage between 
asymptotic giant branch (AGB) stars and planetary nebulae (PNe).  Such objects are expected to be surrounded by an expanding circumstellar envelope of gas and dust ejected during and especially at the end of the AGB phase.  Based on infrared measurements made by the {\it Infrared Astronomical Satellite} ({\it IRAS}), a number of candidates were identified based on their large infrared excesses at 25 $\mu$m, resulting from emission by the cool (150$-$300 K) dust \citep[e.g.,][]{partha86,hri89}.
Follow-up millimeter wavelength observations in some sources revealed the molecular component of the gas \citep{lik89,lik91,omont93} and {\it Hubble Space Telescope} ({\it HST}) observations revealed the envelope in scattered light \citep{ueta00,sahai07,siod08}.

Observations of the central stars of PPNe showed spectral types from late-G to early-B, with the luminosity class of supergiants \citep{suarez06} due to their low surface gravities.
 High-resolutions spectra indicated that they are divided between carbon and oxygen-rich chemistries and are iron-poor compared to the Sun \citep{vanwin03}.
The stars were observed to vary in brightness, and intensive studies were carried out by \citet{ark10,ark11} and  \citet{hri10,hri15a}.
The light variability is complex, with the appearance of varying periods and amplitudes, and can be variously described as somewhat periodic, multi-periodic, or quasi-periodic.  
The maximum amplitudes (peak to peak) of PPNe are not large, $\le$0.7 mag and typically $\le$0.25 mag.
Similar complexities are seen in other post-AGB stars, such as RV Tauri variables \citep{kiss07}, although since they are typically larger-amplitude pulsators than the PPNe, the complexities are not as dominant.
Periods of PPNe have been found to range from 35 to 160 days \citep{hri10,hri15b}, and well-studied ones show what appear to be multiple periods beating upon one another, usually with a secondary period close ($\pm$10 $\%$) to the primary \citep{hri11,ark11,hri15a}.

A combined light and velocity curve pulsation study of two bright PPNe was carried out by \citet{hri13}.  
Examining the relationship between light, color, and radial velocity curves, they found that the phases of the color curves were identical with that of the {\it V} light curve, while the phases of the radial velocity differed by about $-$0.25 phase. 
In this paper, we begin by investigating the periodicity in the radial velocity curves of three additional bright PPNe.  We then go on to examine the relationship between light, color, and radial velocity curves for three PPNe.   These phase relationships can help to constrain the pulsation models for  objects, and the associated periods and amplitudes can, in principle, be used to determine their masses. 

\section{PROGRAM OBJECTS}
\label{object]}

The three program objects, IRAS 17436+5003, 18095+2704, and 19475+3119,  
are among the seven bright ({\it V}$<$10.5) PPNe that we have been monitoring for radial velocity variability \citep{hri17}.
They clearly possess the properties of PPNe.  All have double-peaked spectral energy distributions, with a peak in the visible arising from the (reddened) photosphere and a second peak in the mid-infrared arising from re-radiation from cool (T$\approx$200 K) dust.  
These three possess expanding circumstellar shells of gas and display the spectra of F supergiants.  
Some basic properties of these objects are listed in Table~\ref{object_list}, including effective temperature ({\it T$_{\rm eff}$}) and surface gravity ({\it g}).
They are metal poor ([Fe/H]$<$0)), as befits their evolved age, and oxygen-rich ([C/O]$<$0).
 No evidence of binarity has been found in these three objects \citep{hri17}.

\placetable{object_list}

For each of these objects, we have obtained contemporaneous light and radial velocity curves from 2007$-$2015, and we have also obtained radial velocity curves from 1991$-$1995.  
We recently published a detailed light curve study of these three based on our observations from 1994 to 2012 \citep{hri15a}.  There are also other published light curve studies and for IRAS 17436+5003, a published radial velocity study from the late 1970s \citep{bur80}. 

\section{RADIAL VELOCITY OBSERVATIONS}
\label{velocities}

We have carried out radial velocity observations of these three objects using high-resolution spectra observed with four different telescope-spectrograph-detector systems.
The initial observations were carried out at the Dominion Astrophysical Observatory (DAO) in Victoria, Canada, using the Radial Velocity Spectrometer at the Coud\'{e} focus of the 1.2-m telescope.  
This used a physical mask based on an F star and covered the spectral range 4000$-$4600 \AA.  Most of the data are from 1991 to 1993.  
We will refer to these as the DAO-RVS observations. 
We re-initiated the radial velocity program at the DAO in 2007, using a CCD (DAO-CCD) and covering a smaller spectral region, 4350$-$4500 \AA. 
More details of the DAO observations are given in our earlier papers \citep{hri11,hri13}. 

The study was expanded to include radial velocities observations for all three objects with the the Flemish 1.2-m Mercator Telescope on La Palma and additional observations of IRAS 17436+5003 with the 1.65-m telescope of the Moletai Observatory (Lithuania).  
The Mercator observations began in 2009 and used the HERMES fiber-fed echelle spectrograph \citep{rask11} and covered the spectral range 4770$-$6550 \AA.  
The Moletai Observatory spectra covered the interval 3850$-$6400 \AA~ and used a physical CORAVEL mask based primarily on the solar spectrum.  These Moletai observations were carried out from 2008 to 2014. 

Despite our care in calibrating the observations made with the different telescope-spectrograph-detector systems, we found systematic differences between them for each of the individual stars.  
The values of these systematic differences were determined empirically for the three contemporaneous radial velocity data sets observed from 2007 to 2015, and the details are given in Appendix I.  
Adopting the higher-precision HERMES data as the standard, we found that the differences ranged from 0.5 to 1.45 km~s$^{-1}$ for the three objects.  
We attribute these differences to peculiarities of the PPNe spectra as compared with standard stars, including shocks and outflows, and the differences in the wavelength regions used.  
This is discussed in much more detail in our recent radial velocity search for evidence of binary motion in these three and four other bright PPNe \citep{hri17}.
We adjusted the 2007$-$2015 data from these three telescopes for the empirically-determined differences (offsets), and then combined them into one radial velocity data set for each of these three objects.
The 1991$-$1995 data sets were examined separately.

A summary of the different radial velocity data sets for each object, including the years of observations, the number of observations, the average values, and the offset values, is listed in Table~\ref{data_sets}.  These average values are as measured, without including the empirically-determined offsets. 

\placetable{data_sets}

In Tables~\ref{rvdata_17436}, \ref{rvdata_18095}, and \ref{rvdata_19475} are listed the individual radial velocity measurements for IRAS 17436+5003, 18095+2704, and 19475+3119, respectively, with each of the different telescope-spectrograph-detector systems (DAO-RVS, DAO-CCD, HERMES, CORAVEL). 
Again, these listed values are as measured on the individual systems, without the offsets included.

\placetable{rvdata_17436}

\placetable{rvdata_18095}

\placetable{rvdata_19475}

\section{NEW PHOTOMETRIC OBSERVATIONS}
\label{photo}

New photometric observations of these objects were carried out from 2013 to 2015 at the Valparaiso University Observatory (VUO).  These were obtained with an 0.4 m telescope equipped with an SBIG 6303 CCD camera and standardized to the Johnson {\it B} and {\it V} and Cousins {\it R$_C$} systems.  
Differential photometry was employed, using an aperture of 11$\arcsec$ diameter, as was used in our previous studies.
The number of observations of each are as follows $-$ IRAS 17436+5003: 104 ({\it B}), 153 ({\it V}), 158 ({\it R}); IRAS 18095+2704: 74 ({\it B}), 115 ({\it V}), 116 ({\it R}); IRAS 19475+3119: 83 ({\it B}), 121 ({\it V}), 121 ({\it R}).
An observation consists of a single exposure in a filter per night.   Note that we do not have {\it B} data from 2015.
Typical exposure times ranged from 6 sec for IRAS 17436+5003 to 2 min for IRAS 18095+2704.
The differential standardized photometric values are listed in Tables~\ref{lcdata_17436}, \ref{lcdata_18095}, and \ref{lcdata_19475}.  These complement and were combined with our earlier 2007$-$2012 data sets in the following analyses.  The same comparison stars were used as we used previously (GSC  03518-00926 for IRAS 17436+5003, GSC 02100-00387 for IRAS 18095+2704, and GSC 02669-02709 for IRAS 19475+3119), and their magnitudes are listed in our previous paper \citep{hri15a}. 

\placetable{lcdata_17436}
\placetable{lcdata_18095}
\placetable{lcdata_19475}

\section{PERIOD ANALYSES}
\label{Per-Anal}

Period analyses were carried out using PERIOD04 \citep{lenz05}, which uses a Fourier transform to determine the frequencies in the data.  It is written to easily accommodate multiple periods using sine curve fits.
We adopted their criteria of significance, which is a signal-to-noise ratio (S/N) greater than or equal to 4.0 \citep{bre93}. 
Similar period values were also determined using the generalized Lomb-Scargle periodogram \citep{zech09}.

\section{VARIABILITY STUDY OF IRAS \\17436+5003 (HD 161796)}
\label{17436}

\subsection{Light and Color Curve Study}

Fernie, in a series of papers culminating in 1995 \citep{fern95}, obtained light and color curves of IRAS 17436+5003 from 1980$-$1998\footnote{Data for all of these years, including 1995$-$1998, were obtained in 2012 from the David Dunlap Observatory website, but are no longer available from there.} and more recently \citet{hri15a} carried out light and color curve observations from 1984$-$2012.  These light curves show a cyclical behavior with varying amplitudes that are small in {\it V} ($\le$0.21 mag, peak to peak) and vary over a range of a factor of 3$-$4.  The amplitudes are larger in ({\it B} $\sim$140\%) and smaller in {\it R} ($\sim$90$\%$).  In our recent study, an analysis of these two data sets combined, along with data from 1979$-$1980 by \citet{per81}, revealed a dominant period of 45.15$\pm$0.01 days over this 35 year interval, along with several additional weaker but significant periods.  
The secondary period was close to the primary period, with {\it P$_2$/P$_1$} = 1.06.  The color varied with brightness, being redder when fainter, and the {\it V} light curve showed an overall increase in brightness of $\sim$0.10 mag over this 35 year interval. 

\subsection{Radial Velocity Study}

IRAS 17436+5003 is bright ({\it V}=7.1) and a relatively easy target for high-resolution spectrographic observations.  
The first series of radial velocity measurements were carried out by \citet{bur80}, who observed it with the CORAVEL spectrophotometer at the Geneva Observatory.  They used a mask based upon Arcturus that covered the spectral region $~$3600$-$5200 \AA,~ and they obtained 74 observations during the 1978 and 1979 seasons, with a typical uncertainty of 0.55 km~s$^{-1}$.  From these radial velocities, they determined a period of 54 days.  
We subsequently observed it from 1991$-$1995 with the DAO-RVS (59 observations) and then from 2007$-$2015 with DAO-CCD (121), HERMES (105), and CORAVEL (107).

We began by analyzing the radial velocities in these three different epochs.
Our period analysis of the \citet{bur80} radial velocities yielded a period of 53.4$\pm$0.4 days, but the S/N for this period (3.9) is slightly below the significance level adopted for PERIOD04 analyses (S/N$\ge$4.0).   
However, the period analysis of these radial velocities year by year  yielded significant periods of  50.3$\pm$0.7 days  (1978, 17 observations) and 56.9$\pm$1.7 days (1979, 57).  
An analysis of the DAO-RVS data set from 1991 to 1995 did not yield a significant period,
although an analysis of a two-year interval with the majority of the data, 1991$-$1992 with 34 observations, did yield a significant period of 42.8$\pm$0.4 days. 
The radial velocities from the contemporaneous DAO-CCD, CORAVEL, and HERMES data sets were initially examined individually before they were combined.
An analysis of the DAO-CCD data from 2007-2015 yielded a significant period of 47.7$\pm$0.1 days.  
An analysis of the HERMES data found a most-likely period of 51.8 days. 
The CORAVEL data yielded two marginally significant periods, 35.3$\pm$0.1 and 46.6$\pm$0.1 days. 
These three data sets were then combined, using empirically-determined offsets to correct for the measured systematic differences between them (see Appendix I).
The full range of velocities (peak-to-peak) is 13 km~s$^{-1}$.
The period analysis of the combined 2007$-$2015 data set led to a value of 39.4 days, but it was not formally significant (S/N=3.4).
The results of the periodigram analyses of these various subsets of the radial velocity data are summarized in Table~\ref{rv_results}.  
In this table are listed the systemic velocities ({\it A$_0$}), periods ({\it P}), amplitudes ({\it A}), and phases ({\it $\phi$}) for the indicated observing years.
One sees that the recent observations analyzed separately indicate periodicities in the data, but the values do not all agree.
The results of these analyses highlight the impact of the complexities in the velocity curve on the analysis of limited data sets for such an object with a multi-periodic or quasi-periodic, semi-regular variability.

\placetable{rv_results}

\subsection{Contemporaneous Light, Color, and Velocity Curve Study}

\begin{figure}\figurenum{1}\epsscale{1.8}
\plotone{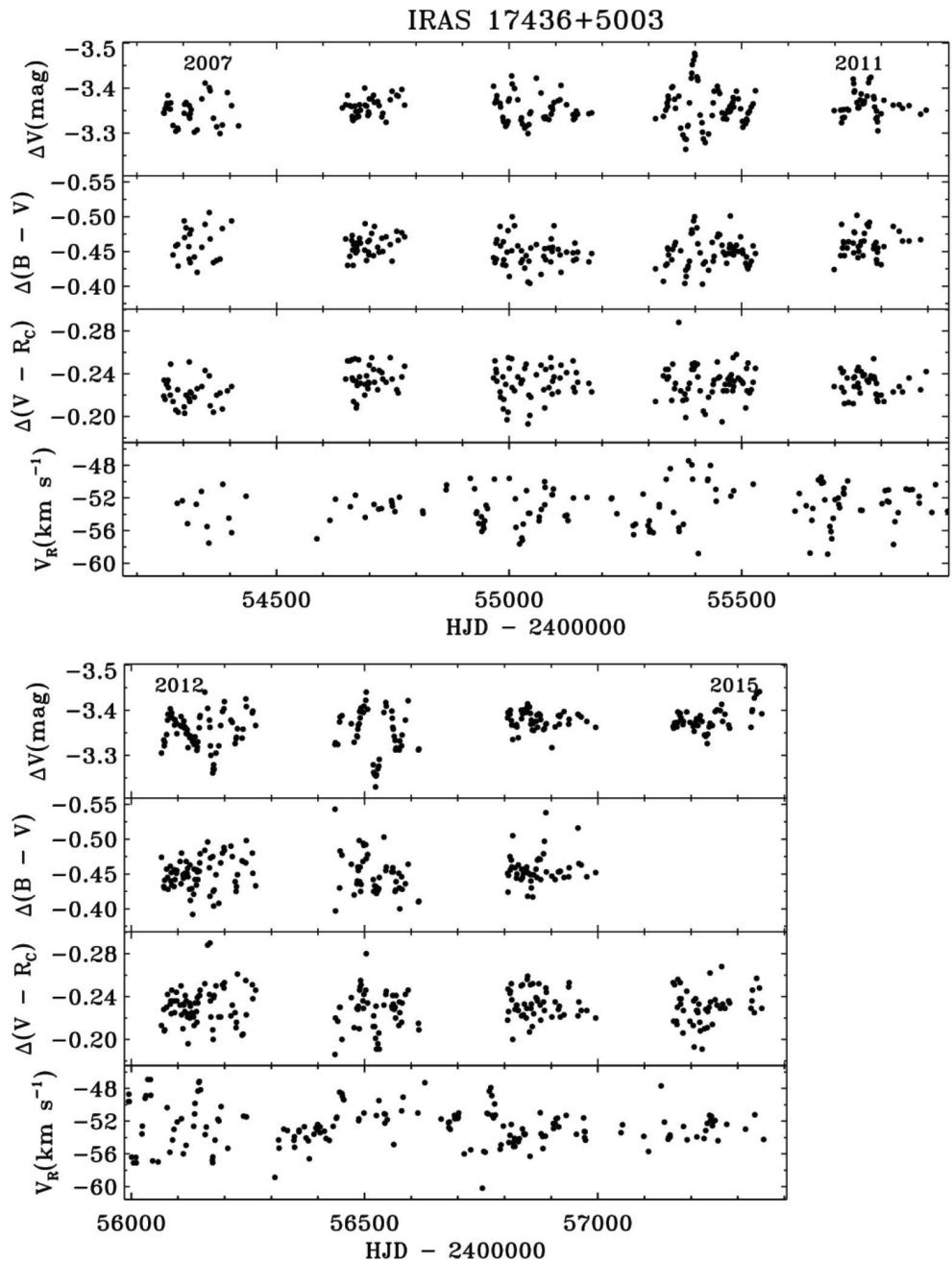}
\caption{IRAS 17436+5003: 2007$-$2015.  Photometry from 2007$-$2012 from \citet{hri15a} and from 2013$-$2015 from this study; radial velocities from our DAO-CCD, CORAVEL, and HERMES observations. 
Sample years of the observations are indicated in the plots.
({\it B$-$V} and {\it V$-$R$_C$} are in units of mag in this figure and the following ones.)
\label{17436-1}}
\epsscale{1.0}
\end{figure}

In Figure~\ref{17436-1} are shown the contemporaneous light ({\it V}), color ({\it B$-$V} \& {\it V$-$R$_C$}), and radial velocity ({\it V$_R$}) curves of IRAS 17436+5003 from 2007$-$2015.
These are based on the light and color observations by \citet{hri15a} and our more recent 2013$-$2015 observations.  The radial velocities consist of the combined results from DAO-CCD (2007$-$2015), CORAVEL (2008$-$2014), and HERMES (2009$-$2015), adjusted with the empirically-determined offsets. 

\placefigure{17436-1}

\begin{figure}\figurenum{2}\epsscale{0.65}
\plotone{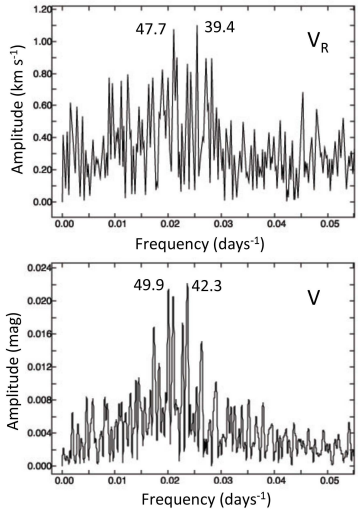}
\caption{Frequency Spectrum for IRAS 17436+5003 (2007$-$2015) $-$ top: radial velocity data, with peak corresponding to 39.4 and 47.7 days (neither significant); bottom: {\it V} light curve data, with a peaks at 42.3 and 49.9 days (significant).  
\label{17436_FreqSpec}}
\epsscale{1.0}
\end{figure}

Visual inspection of the light curves show periodic or quasi-periodic variation in some of these seasons (2007, 2009, 2010, 2012, 2013) but not in others.  
Those with the periodic or quasi-periodic variations were generally those with the largest peak-to-peak seasonal variations.  The overall seasonal variations ranged in {\it V} from 0.08$-$0.22 mag.
A formal period analysis of the 2007$-$2015 {\it V} light curve revealed two periods in the data, 42.3 and 49.9 days, with similar amplitudes.  
The same values were found for the {\it B}, {\it R$_C$}, and ({\it B$-$V}) curves.  No periodicity was found in the overall ({\it V$-$R$_C$}) curve, which has the smallest amplitude variations (0.06 mag peak-to-peak).
Similar values were found when we adjusted the data to remove slight differences in the seasonal mean levels.
The frequency spectrum for the {\it V} light curve from 2007 to 2015 and also for the radial velocity over that same time interval are shown in Figure~\ref{17436_FreqSpec}.
Analyzing the {\it V}  light curve year by year resulted in significant periods in 2007 (43.8 days), 2009 (48.8 days), 2010 (45.6 days),  and 2013 (47.8 days),
with approximately similar values found in the {\it B}, {\it R$_C$}, and ({\it B$-$V}) curves.
As noted previously, no significant period is found in the combined radial velocity curve over this 2007$-$2015 interval. 
A year-by-year analysis of the radial velocity data revealed significant periods in 2007 (45.9 days), 2010 (50.0 days), 2012 (53.0 days), and a marginally significant period in 2011 (48.6 days).  
Thus significant periods are present in some but not all of the individual years for both the light and velocity curves.
For the two years that do have periods in both, 2007 and 2010, the period values are close but not identical, with those found in the radial velocities $\sim$7~$\%$ larger.  

\placefigure{17436_FreqSpec}

\begin{figure}\figurenum{3}\epsscale{0.7}
\plotone{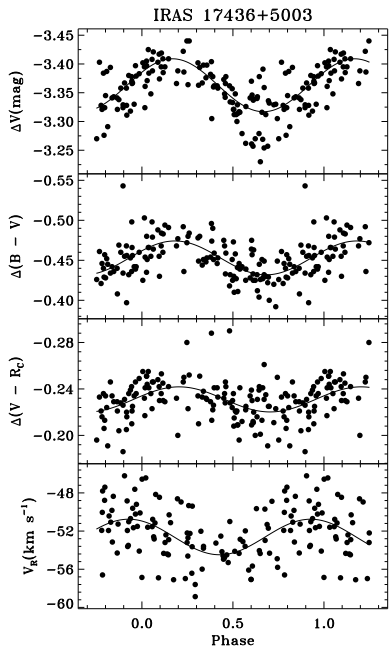}
\caption{IRAS 17436+5003: phased plots for the 2012$-$2013 observations, with P=49.52 days and sine curve fits.
\label{17436_phase_12-13}}
\epsscale{1.0}
\end{figure}

For the purposes of comparing the light, color, and velocity curves, we began by examining the data in two-year rather than one-year intervals, in an effort to increase the number of data points and obtain a more robust result. 
This was done for the years showing the most cyclical behavior, 2009$-$2010 and 2012$-$2013. 
Investigating in detail the 2012$-$2013 data using the period derived from the {\it V} light curve ({\it P}=49.52 days) and an arbitrary epoch of JD 2,455,600.00 yielded for the respective data sets the following semi-amplitudes ({\it A}) and phases ({\it $\phi$}) for sine curve fits $-$ {\it V}: 0.046 mag and 0.58, ({\it B$-$V}): 0.021 mag and 0.57, ({\it V$-$R$_C$}): 0.011 mag and 0.55, {\it V$_R$}: 1.80 km~s$^{-1}$ and 0.32, with an uncertainty in phase of $\pm$0.02$-$0.03.  These curves are shown in Figure~\ref{17436_phase_12-13} and the sine curve parameters are listed in Table~\ref{fit_par}.  
The variation in the amplitudes seen in the overall light curve (Fig.~\ref{17436-1}) is shown clearly in the phased {\it V} light curve.
The standard deviations from the sine curves are as follows: $\sigma$({\it V}) = 0.031 mag, $\sigma$({\it B$-$V}) = 0.021 mag,  $\sigma$({\it V$-$R$_C$}) = 0.016 mag, and $\sigma$({\it V$_R$}) = 2.42 km s$^{-1}$.
The light and color curves are in phase, with the radial velocity curve a quarter of a cycle out of phase.
A similar investigation of the 2009$-$2010 data using the period derived from the {\it V} light curve ({\it P}=48.05 days) did not give a good fit to either of the color or radial velocity curves.

\placefigure{17436_phase_12-13}

\placetable{fit_par}

We next investigated the 2007 and 2010 data in individual years.  The 2007 {\it V} light curve has a period of 43.8 days, but there was not a consistent period that could be applied to all of the data of that year.  However, the 2010 data are more consistent, with a similar period of 45.6$\pm$0.1 days found in the {\it V}, {\it B}, {\it R$_C$}, and ({\it B$-$V}).  When we fit the color and velocity curves to the period found in the {\it V} light curve, 45.62 days, we found the following semi-amplitudes and phases for sine curve fits $-$ {\it V}: 0.052 mag and 0.18, ({\it B$-$V}): 0.019 mag and 0.15, ({\it V$-$R$_C$}): 0.009 mag and 0.19, {\it V$_R$}: 2.88 km~s$^{-1}$ and 0.98, with an uncertainty in phase of $\pm$0.02$-$0.04.  

\begin{figure}\figurenum{4}\epsscale{1.4}
\plotone{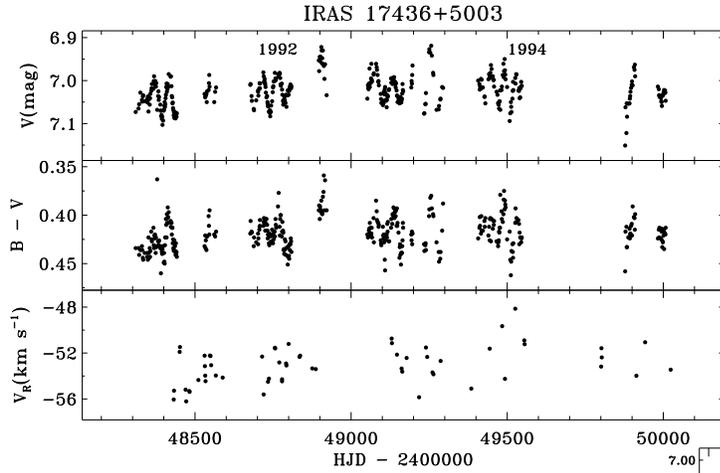}
\caption{IRAS 17436+5003: 1991$-$1995.  Photometry from \cite{fern93,fern95}; radial velocities from DAO-RVS.  Sample years of the observations are indicated in the plot.  
\label{17436-3}}
\epsscale{1.0}
\end{figure}

\begin{figure}\figurenum{5}\epsscale{0.7}
\plotone{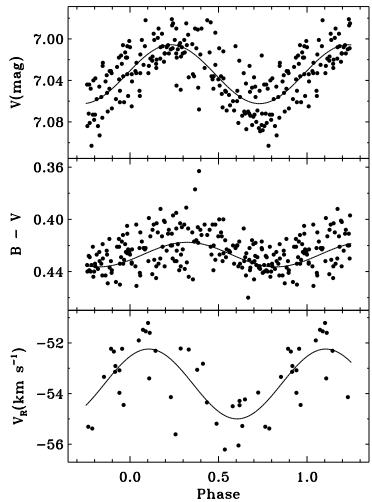}
\caption{IRAS 17436+5003: phased plots for the 1991$-$1992 observations, with P=43.34 days and sine curve fits.  
\label{17436_phase_91-92}}
\epsscale{1.0}
\end{figure}

There are two additional sets of contemporaneous data that we examined.  
The first of these is our DAO-RVS radial velocities and the light and color curves of \citet{fern93,fern95} from 1991$-$1995; these data are shown in Figure~\ref{17436-3}.  
Since the 1991$-$1992 radial velocities possess a periodicity, we carried out an analysis for these years.  
Examination of the 1991$-$1992 light curves also reveal cyclical behaviour over most of this time.  
The photometry in the later part of 1992, however, shows a relatively large increase in brightness and blueness.
Since we are not able to model these increases and since we did not have radial velocity observations during this time to see if there was an impact on the velocity measurements, we have excluded these photometric data from our periodogram analysis.
Analyzing the 1991$-$1992 measurements, excluding the photometry at the end of 1992, results in {\it P(V)} = 43.3 days, {\it P}({\it B$-$V}) = 44.2 days, and {\it P}({\it V$_R$}) = 42.8 days.
An analysis of the light, color, and velocity curves with the period fixed at that of the {\it V} light curve resulted in the following semi-amplitudes and phases $-$ {\it V}: 0.028 mag and 0.52, ({\it B$-$V}): 0.009 mag and 0.43, {\it V$_R$}: 1.39 km~s$^{-1}$ and 0.14, with an uncertainty in phase of $\pm$0.01$-$0.03. 
 These phase curves are shown in Figure~\ref{17436_phase_91-92}.  
 The fits to the sine curves are good, with standard deviations of $\sigma$({\it V}) = 0.019 mag, $\sigma$({\it B$-$V}) = 0.013 mag, and $\sigma$({\it V$_R$}) = 0.97 km s$^{-1}$.

\placefigure{17436-3}

\placefigure{17436_phase_91-92}

\begin{figure}\figurenum{6}\epsscale{1.0}
\plotone{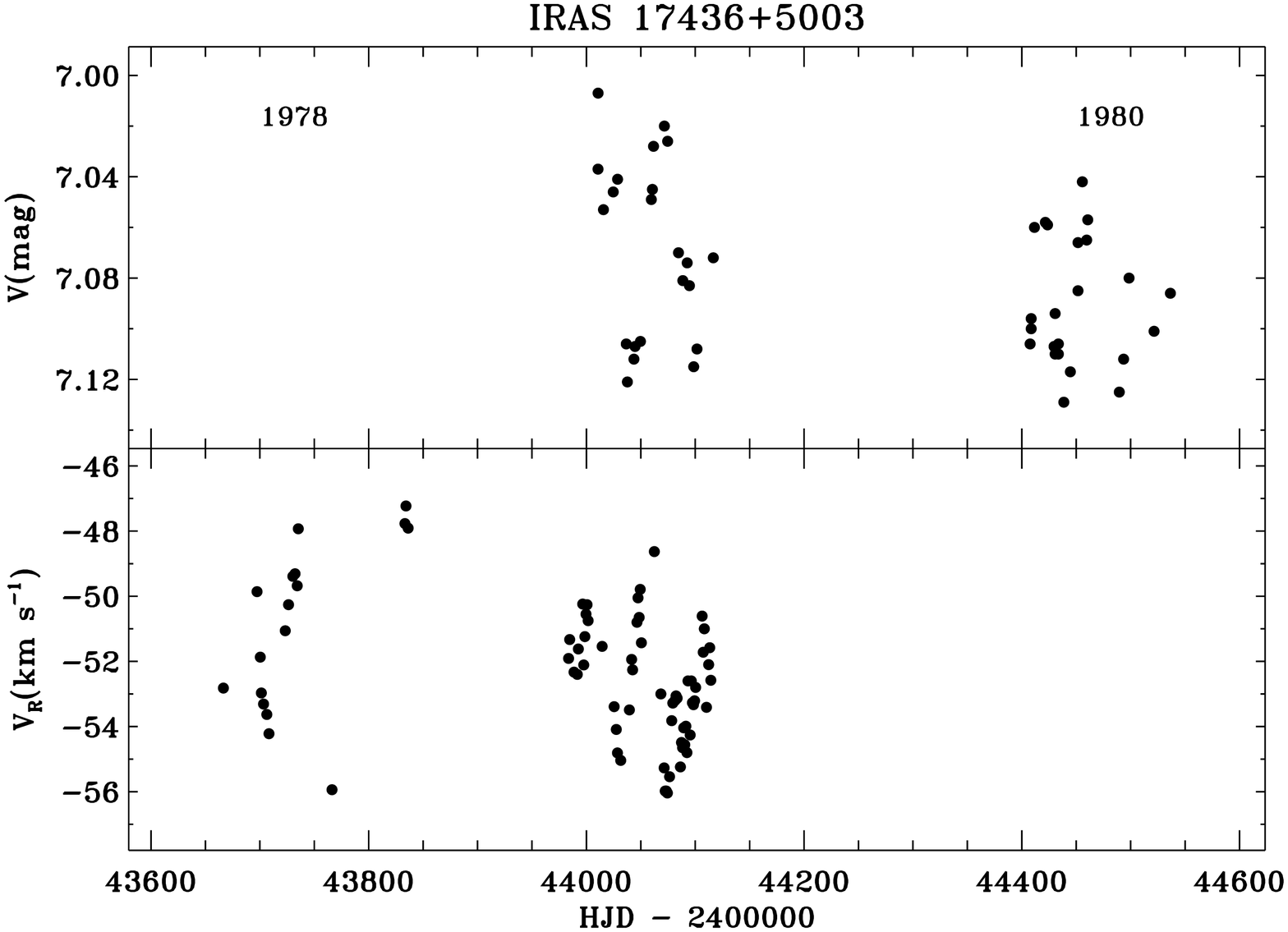}
\caption{IRAS 17436+5003: 1978$-$1980.  Photometry from \citet{per81} and \cite{fern83}; radial velocities from \citet{bur80}.  Sample years of the observations are indicated in the plot.
\label{17436-4}}
\epsscale{1.0}
\end{figure}

The other set of contemporaneous velocity and photometry data are the radial velocity measurements of \citet{bur80} from 1978$-$1979 and the {\it V} light curve measurements of \citet{per81} and \cite{fern83} from 1979$-$1980, shown in Figure~\ref{17436-4}.  We investigated the contemporaneous 1979 observations.   These resulted in {\it P(V)} = 59.1 $\pm $1.9 days and {\it P}({\it V$_R$}) = 56.9 $\pm$1.7 days.  When we fixed the period at that of the {\it V} light curve data, we find the following semi-amplitudes and phases $-$ {\it V}: 0.043 mag and 0.68, {\it V$_R$}: 1.75 km~s$^{-1}$ and 0.49, with an uncertainty in phase of $\pm$0.02.  
The results of the periodigram analyses of these various additional radial velocity subsets are listed in Table~\ref{rv_results} and the results for the light, color, and velocity curve fits are listed in Table~\ref{fit_par}.

\placefigure{17436-4}

\section{VARIABILITY STUDY OF IRAS \\18095+2704 (HD 335675)}
\label{18095}

\subsection{Light and Color Curve Study}

\citet{ark10} and more recently \citet{hri15a} have carried out photometric studies of IRAS 18095+2704.  In the latter study, we presented our photometric light curves from 1994$-$2012, which showed a cyclical variation in light and color, along with a continual slow increase in brightness.  These data were combined with those of \citet{ark10}, beginning in 1993, into a long-term period study.  A primary period of 113.2$\pm$ 0.1 day was determined from an analysis of the entire data set, while subsets over shorter intervals of time resulted in period values of 98$-$114 day.  
While cyclical, these light curves showed a large variation in amplitude, from 0.05 to 0.14 mag in {\it V} (peak to peak).  
The amplitudes were systematically larger in {\it B} and smaller in {\it R}.
This cyclical variation is superimposed on an approximately monotonic increase in brightness of 0.40 mag ({\it V}) over 20 years.
Multiple periods were found in these analyses, with secondary periods relatively close to the primary period and a ratio {\it P$_2$/P$_1$} of 0.86.  The color varied with brightness, being redder when fainter.  

\subsection{Radial Velocity Study}

Visual inspection of the DAO-RVS, DAO-CCD, and HERMES radial velocity data sets revealed cyclical variations in most of the seasons that had 10 or more observations.
The different radial velocity data sets were investigated individually.
A period search of the 1991$-$1995 DAO-RVS velocities revealed a period of 109.2$\pm$0.6 day.
The 2007$-$2015 DAO-CCD velocities suggested a period of 103.5$\pm$0.4 days, but below the significance criteria. 
An identical period of 103.5$\pm$0.5 days was also found in the the 2009-2015 HERMES data, but again at a level slightly below the significance criteria.  
The analysis of the combined data from 2007 to 2015 resulted in a period of 103.5$\pm$0.3 day that was significant and a velocity semi-amplitude of 1.08 km~s$^{-1}$.  
The combined DAO-CCD and HERMES radial velocity curve displays a range of 10 kms$^{-1}$ peak to peak.
The results of the period analyses of these radial velocity curves are listed in Table~\ref{rv_results}.

\subsection{Contemporaneous Light, Color, and Velocity Curve Study}

\begin{figure}\figurenum{7}\epsscale{1.8}
\plotone{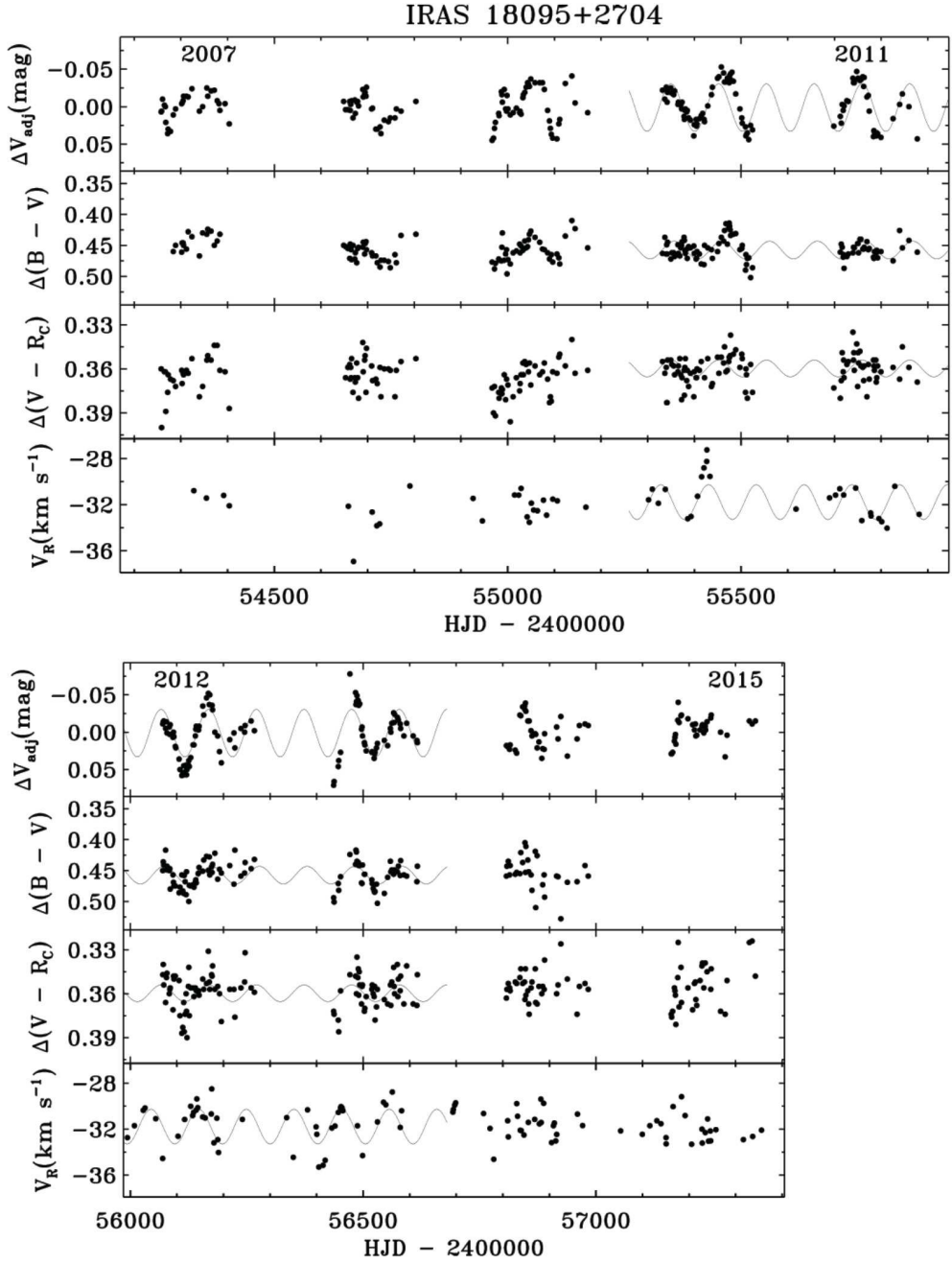}
\caption{IRAS 18095+2704: 2007$-$2015.  Photometry from 2007$-$2012 from \citet{hri15a} and from 2013$-$2015 from this study,  with the brightening trend removed; radial velocities from our DAO-CCD and HERMES observations. 
The solid lines are sine curve fits determined from the analysis using a period of 102.3 days derived from the fit to the 2010$-$2013 {\it V} light curve.  Sample years of the observations are indicated in the plots.
\label{18095_obs}}
\epsscale{1.0}
\end{figure}

\begin{figure}\figurenum{8}\epsscale{0.7}
\plotone{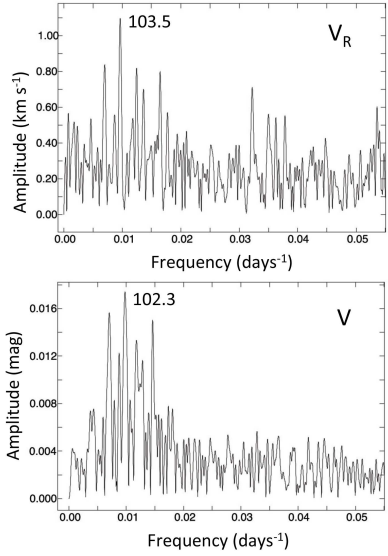}
\caption{Frequency Spectrum for IRAS 18095+2704 (2007$-$2015) $-$ top: radial velocity data, with peak corresponding to 103.5 days;  bottom: {\it V} light curve data, with a peak at 102.3 days.  Both are significant.
\label{18095_FreqSpec}}
\epsscale{1.0}
\end{figure}

In Figure~\ref{18095_obs} are displayed the contemporaneous light, color, and velocity curves from 
2007$-$2015, the years for which we have data sets in common.  
The light and color curves are from VUO and the radial velocity curves from DAO-CCD and HERMES. 
The brightness increase in the light curve was removed by fitting it with a low-order polynomial.
Cyclical variations can be seen in the {\it V} light curves from each year, with peak-to-peak amplitudes ranging from 0.06 (2007, 2008) to 0.15 (2013).  Visual inspection of the {\it V} light curve shows that the periods of the variation range from 90$-$130 days.  However in some seasons there appears to be a shallower minimum approximately half way between the deeper minima; we ascribe this to the multi-periodic character of the light variations.
A period analysis of the {\it V} light curve from 2007$-$2015 yielded a dominant period of 102.3$\pm$0.2 days.
This frequency spectrum is shown in Figure~\ref{18095_FreqSpec}, along with that of the combined radial velocities.
Additional significant periods were found in the {\it V} data, and when determined simultaneously for the first three periods, yielded {\it P$_1$} = 103.6, {\it P$_2$} = 77.8, and {\it P$_3$} = 137.0 days.
There are also ASAS-SN photometric data\footnote{All-Sky Automated Survey for Supernovae; https://asas-sn.osu.edu} available for this object that cover the years 2013$-$2017 \citep{koch17}.  The ASAS-SN data set includes more points with higher frequency but somewhat lower precision, and they show even more detailed and complex structure in the light curve.  A period analysis of this data set yields a significant period of 105.4 days. 
Thus the more recent data ($\ge$ 2007) indicate a shorter period than that found for the earlier 1994$-$2012 data.  

\placefigure{18095_obs}

\placefigure{18095_FreqSpec}

\begin{figure}\figurenum{9}\epsscale{0.7}
\plotone{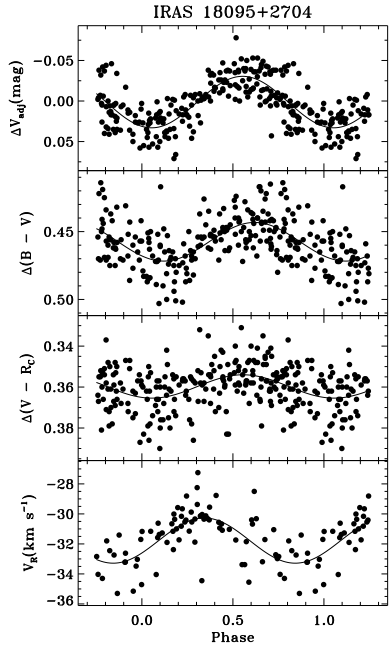}
\caption{IRAS 18095+2704: phased plots for the 2010$-$2013 observations with {\it P}=102.3 days  and sine curve fits.  
\label{18095_phase_10-13}}
\epsscale{1.0}
\end{figure}

To investigate the phase relationship between our light, color, and radial velocity curves, we compared the results of a period analysis of our 2010$-$2013 data sets.  These four seasons have the largest light curve amplitudes and each shows a clear cyclical variation with a period of $\sim$100 days.
A dominant period of 102.3$\pm$0.2 days was determined from the {\it V} light curve, with similar values found in the {\it B} and {\it R} light curves and the {\it B$-$V} color curve.  
Using the period derived from the {\it V} light curve and an arbitrary epoch of JD 2,455,600.00, together with a sine curve fit, yielded the following semi-amplitudes and phases $-$ {\it V}: 0.032 mag and 0.20, {\it B}: 0.044 mag and 0.18, {\it R}: 0.026 mag and 0.20, {\it B$-$V}: 0.014 mag and 0.13, {\it V$-$R}: 0.006 mag and 0.18, and V$_R$: 1.51 km~s$^{-1}$ and 0.91, with an uncertainty in phase of $\pm$0.01.  These curves are superimposed on the data in Figure~\ref{18095_obs}.
The sine curve fits the cyclical data of the {\it V} light curve for 2010$-$2013 reasonably well, although the fact that the sine curve is of constant amplitude and we have used only the dominant period and not secondary periods results in clear intervals of systematic deviations.  Nevertheless, it is a good representation of the average {\it V} variation over these years.  The fits to the ({\it B$-$V}) and  {\it V$_R$} are not as good but do serve to represent the data.  For the {\it V$-$R} data, the variation is on the order of the precision and a good fit to a sine curve is not to be expected.
These fit parameters are listed in Table~\ref{fit_par}. 
In Figure~\ref{18095_phase_10-13} are shown the phase plots for the contemporaneous light, color, and radial velocity curves for 2010$-$2013. 
The fits are reasonably good, given the observed variations in the amplitudes,
with standard deviations of $\sigma$({\it V}) = 0.019 mag, $\sigma$({\it B$-$V}) = 0.015 mag, $\sigma$({\it V$-$R$_C$}) = 0.010 mag, and $\sigma$({\it V$_R$}) = 1.27 km s$^{-1}$.
(Note that the analysis of the radial velocities over this 2010$-$2013 time interval resulted in a period of 103.4$\pm$0.6 days.  However, the fit to the radial velocities with this period is only marginally better, with $\sigma$({\it V$_R$}) = 1.24 km s$^{-1}$.)
One sees that the light and color curves are essentially in phase, with the {\it B$-$V} color curve peaking slightly after the {\it V} light curve.  
The radial velocity curve, by contrast, is approximately a quarter of a cycle 
out of phase with the light curve.  

\placefigure{18095_phase_10-13}

\section{VARIABILITY STUDY OF IRAS \\19475$+$3119 (HD 331319)}
\label{19475}

\subsection{Light and Color Curve Study}
Light curve studies of IRAS 19475+3119 were carried out by \citet{ark06} from 2002$-$2005 and more recently by \citet{hri15a} from 1994$-$2012.   We found the light curve to be multi-periodic or quasi-periodic, with period values, of  41, 39, and 48 days (in order of significance).  The light curves show a varying amplitude and small changes in the yearly mean brightness levels. The amplitudes varied from 0.07 to 0.19 mag ({\it V}), and were larger in {\it B} and smaller in {\it R$_C$}.  The object was redder when fainter.  We have extended these light curves in time with the inclusion of our 2013$-$2015 data from the VUO.

\subsection{Radial Velocity Study}

Radial velocity observations of IRAS 19475+3119 were obtained with the DAO-RVS, DAO-CCD, and HERMES systems, and visual inspection of the data showed clear variability.   
The DAO-RVS measurements, from 1991$-$1995, resulted in a significant period of 47.1$\pm$ 0.1 days.  
No significant period was found in the analysis of the entire DAO-CCD data set, from 2007$-$2015.  The analysis of the HERMES 2009$-$2015 data resulted in two significant periods, P$_1$ = 38.0$\pm$0.1 days and P$_2$ = 33.3$\pm$0.1 days.  
An analysis of the combined DAO-CCD  and HERMES data set resulted in a period of P$_1$ = 37.1$\pm$0.1 days and a semi-amplitude of 2.10 km-s$^{-1}$, with a total range of 16 km-s$^{-1}$. 
These radial velocity results are listed in Table~\ref{rv_results}.

\subsection{Contemporaneous Light, Color, and Velocity Curve Study}

\begin{figure}\figurenum{10}\epsscale{1.8}
\plotone{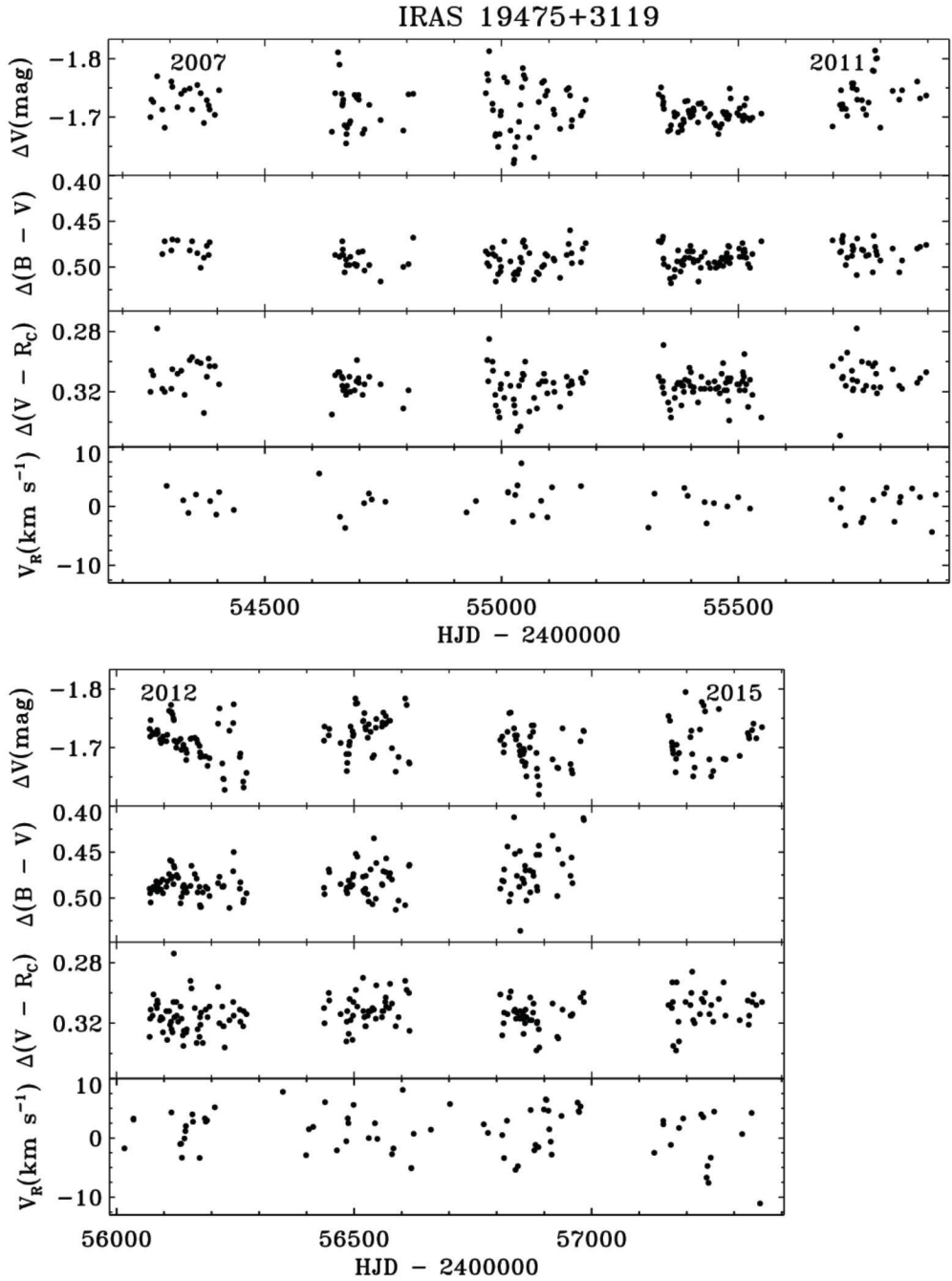}
\caption{IRAS 19475+3119: Photometry from 2007$-$2012 from \citet{hri15a} and from 2013$-$2015 from this study; radial velocities from DAO-CCD and HERMES observations.   Sample years of the observations are indicated in the plots.
\label{19475_obs}}
\epsscale{1.0}
\end{figure}

\begin{figure}\figurenum{11}\epsscale{0.7}
\plotone{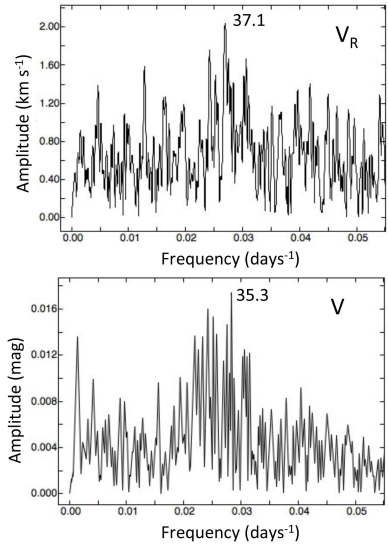}
\caption{Frequency Spectrum for IRAS 19475+3119 $-$ top: radial velocity data, with peak corresponding to 37.1 days; bottom: {\it V} light curve data, with a peak at 35.3 days.  Both are significant.  
\label{19475_FreqSpec}}
\epsscale{1.0}
\end{figure}

In Figure~\ref{19475_obs} is shown the contemporaneous light, color, and radial velocity curves from 2007$-$2015, with the light and color curves based on VUO data and the radial velocity based on the combined HERMES and DAO-CCD observations. 
One can see both the large changes in the seasonal amplitudes and the changes in the mean seasonal brightness levels of the light curve.
A period analysis of the observed {\it V} light curve from 2007$-$2015 resulted in periods of 35.3 and 41.2 days, with similar periods found in the {\it B} and {\it R$_C$} light curves.  
An attempt was made to investigate a periodicity in the longer-term variations in the mean brightness level but none was found.  
We therefore normalized the light curve to the mean seasonal brightness values and again carried out a period analysis of the entire {\it V} light curve.  
This resulted in an identical primary period, P$_1$ = 35.3 days, but a slightly longer secondary period, P$_2$ = 43.1 days. 
The frequency spectrum for the {\it V} light curve, along with that of the radial velocity curve, is shown in Figure~\ref{19475_FreqSpec}.
These photometric periods are similar to those found in the previously published light curve study \citep{hri15a} and the analysis of the 2007$-$2015 radial velocity curves discussed above.
The attempt to compare the light, color, and radial velocity curves was complicated by the multi-periodic nature and lack of a single dominant period throughout the data.  Therefore we examined the individual seasonal light curves to determine the years in which there was a single dominant period in the data.  
Based on the {\it V} light curve, the seasons with significant dominant periods and larger amplitudes were 2008 (37.9 days), 2009 (41.8 days), and 2015 (35.0 days).
Similar, but not identical periods were found for the {\it B} and {\it R$_C$} light curves and for 2009, the ({\it B$-$V}) and ({\it V$-$R$_C$}) color curves.

\placefigure{19475_obs}

\placefigure{19475_FreqSpec}

We also examined the radial velocity data year by year, noting that they have many fewer data points, ranging from 7 to 24 per year.  In most years, no significant periods were found.  The years with a single dominant period are 2012: {\it P} = 41.6 days (17 data points) and 2014: {\it P} = 37.1 days (24 data points),  with 2015 having a significant but less dominant period of 35.3 days (14 data points).

\begin{figure}\figurenum{12}\epsscale{0.7}
\plotone{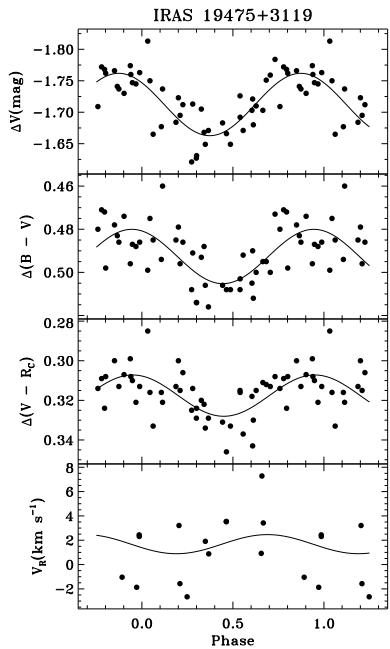}
\caption{IRAS 19475+3119: phased plots for the 2009 observations with {\it P}=41.84 days  and sine curve fits.  The fit is not good to the radial velocity data.
\label{19475_phase_09}}
\epsscale{1.0}
\end{figure}

To compare the contemporaneous light, color, and radial velocity curves, we analyzed the 2009 and 2015 data, which have dominant periods in the {\it V} light curves and also in either the color or radial velocity curves.  
For the 2009 season, with the fixed value of {\it P} = 41.84 days based on the photometric period and an epoch of JD 2,455,600.00, we found semi-amplitudes and phases as follows $-$ {\it V}: 0.050 mag and 0.88, {\it B$-$V}: 0.013 mag and 0.80, {\it V$-$R$_C$}: 0.010 mag and 0.80.  
These parameters resulted in good fits to the light and color curves, with standard deviations of $\sigma$({\it V}) = 0.028 mag, $\sigma$({\it B$-$V}) = 0.010 mag,  and $\sigma$({\it V$-$R$_C$}) = 0.010 mag. 
The resulting V$_R$ curve, however, with semi-amplitude of 2.16 km~s$^{-1}$, phase of 0.64, and standard deviation of 2.14, did not yield a good sine curve fit.  
These sine curve fits are shown in Figure~\ref{19475_phase_09}.

We also compared the 2015 {\it V} light curve and radial velocity curve, since there were good fits for each of them analyzed independently and they were close in value (35.0 and 35.3).  With a fixed value of {\it P} = 35.04 days derived from the {\it V} light curve, we found the following $-$ {\it V}: 0.041 mag and 0.18 and for {\it V}$_R$:  5.10 km~s$^{-1}$ and 0.83.
This yielded a very good fit for each.  These fit results are listed in Table~\ref{fit_par}.

\placefigure{19475_phase_09}

\section{DISCUSSION}

A primary goal of this study was to examine the relationship between the pulsation phases of the light, color, and radial velocity curves for PPNe.  
This phase comparison was determined on certain seasons in which a dominant period was found.
For each of these three objects, the light, color, and velocity curves were fitted well by a sine curve with a period determined from the {\it V} light curve over a limited range of one to four years.  The semi-amplitudes were not large, approximately 0.03$-$0.05 mag for {\it V}, 0.01$-$0.02 mag for ({\it B$-$V)}, and 1.5$-$5 km~s$^{-1}$ for {\it V$_R$}.
In Table~\ref{data_phases} is listed the phase of minimum for each of the light, color, and radial velocity curves tabulated in Table~\ref{fit_par}, along with the phase difference between the color and {\it V} light curve and between the radial velocity and the {\it V} light curve.
For all cases, it was found that the light and color curves were nearly in phase (had similar phase parameters) and the velocity curve was approximately one-quarter cycle out of phase.  This is the same phase relation that had been found previously for the PPNe IRAS 22272+5435 and 22223+4327 \citep{hri13}, and these results are also listed for comparison in Table~\ref{data_phases}.  They had long periods (132, 88 days) and somewhat larger light curve amplitudes, especially IRAS 22272+5435, and consequently somewhat more precise phases.  If we examine these results for all five PPNe in detail, we see that the ({\it B$-$V}) color may peak slightly ($\sim$0.05) after the maximum phase of the {\it V} light curve and the ({\it V$-$R$_C$}) at the maximum phase or slightly ($\sim$0.02) after.  
The radial velocity phases are most accurate for the longer-period objects and they clearly show a quarter-cycle ($-$0.25) difference from the light curve.
These pulsation properties are interpreted to imply that the star is brightest when smallest, with the temperature reaching its maximum value as the star is beginning to expand.

\placetable{data_phases}

These five PPNe have temperatures that place them in the instability strip for evolved stars ({\it T$_{\rm eff}$} = 6500 K and 5750 K, respectively for IRAS 22223+4327 and 22272+5435 \citep{vanwin00, red02}).  Thus we expect that they pulsate due to the $\kappa$ mechanism, with the opacity of the star modulated by the helium partial ionization zone. 
 As remarked upon previously \citep{hri13}, the phasing of these PPNe light and velocity curves is very different from those of classical Cepheids, where they differ by half a cycle.  
Classical Cepheids are not brightest when smallest, but when their photosphere is expanding from minimum size at maximum speed and at approximately average size.  
This phase lag is attributed to the modulating effects of the changing location of the hydrogen partial ionization zone and its associated opacity effects on the energy flux \citep{ostlie07}.  This difference is likely due to the response of the much lower-mass envelope found in these lower mass ($\sim$0.6 M$_{\sun}$) PPNe as compared to the higher-mass classical Cepheids ($>$3 M$_\sun$).

A more appropriate phase comparison is with RV Tauri variables, which are also thought to be post-AGB objects, although not necessarily (and not likely) objects that will evolve into PNe.
These objects are defined by a characteristic pattern of an alternating deeper minimum followed by a shallower one.  Photoelectric light curves are available for some of these for four decades, and some, like those of AC Her, are relatively stable.   
They typically have spectral types ranging from mid-F to K and periods 
ranging from 30 to 150 days between deeper minima ranging (and thus two minima per cycle).  Recently, high precision radial velocity curves have become available for a number of RV Tauri variables, and they reveal not only the pulsational motion of the star but also its orbital motion in a binary system \citep[][and references therein]{manick17}.  Given that these objects have been well studied both photometrically and spectroscopically, it is surprising that there is not a study analogous to ours that explicitly compares the phasing of the two.  
A comprehensive photometric and spectroscopic study of 11 RV Tauri variables was carried out by \citet{poll96,poll97}, but unfortunately there was not enough contemporaneous coverage to accurately compare the light and velocity phasing.
For AC Her, we were able to find enough data to allow a general comparison of its light and velocity curves.  
\citet{gill90} obtained 45 radial velocity observations over a 108 day interval and accurately delineated the radial velocity behavior through 1.4 cycles of its 75-day pulsation period.  The radial velocity curve also has two minima per cycle, with one much shallower than the other.   They also included a schematic drawing of the visual light curve based on observations by the AAVSO light curve observations, in which they claim that the accuracy of the minima is 5 days (0.07 {\it P}).  
Comparing these contemporaneous light and velocity curves reveals that deeper velocity minimum occurs approximately one-quarter of a cycle before the deeper light minimum.  
At our request, R. Manick kindly examined the light and velocity data for the two objects in their sample \citep{manick17} that had contemporaneous observations, IRAS 17038$-$4815 and HP Lyr.  For IRAS 17038$-$4815 (P=76 days), the radial velocity differs by about a quarter of a cycle ($-$0.25 phase).
For HP Lyr, the comparison is ambiguous, as it is one of the few RV Tauri variables which does not appear to show alternating depths of minima, and thus it is uncertain if the period is 68 days or twice that value.  If it is the former, then the radial velocity is a half cycle out of phase and if it is the latter, then the radial velocity is a quarter of a cycle out of phase.
Thus in general, for the two or three cases that we can compare, the phase difference between the light and velocity curves of RV Tauri variables appears to agree with what we have found for our sample of five PPNe studied thus far. 

Regarding the phasing between the light and color curves, for the three RV Tauri variables studied by \citet{zsol90} over several years,  AC Her, R Sge, and V Vul, it was found that their ({\it B$-$V}) color curves reach minimum and maximum $\sim$0.08 {\it P} before the {\it V} light curves.  
This is in contrast to what we have found for the PPNe, where the color curves are in phase with or reach their extrema slightly after the light curves.

These PPNe pulsations are complex, as is seen in the non-repetitive nature of the amplitudes, the multiple periods, and the scatter in the light and velocity curves.  
Similar complex pulsations are seen in the light curves of RV Tauri and other post-AGB objects \citep{kiss07,kiss17}, although in PPNe, which typically have smaller amplitude light variations, they appear relatively more pronounced.
These effects presumably arise from the non-linear nature of the pulsations.  They also manifest themselves in line profile variations, including line splitting.  These complexities have been documented in the spectra of  PPNe by \citet{leb96}, \citet{zacs09, zacs16}, and \citet{kloch14}.

There have been very few modeling efforts for pulsation in cool post-AGB stars, and the results do not agree well with  these and previous PPNe observations.  \citet{fok01} computed non-linear models for post-AGB stars in an attempt to model the PPNe IRAS 07134+1005 (HD 56126).  This is an F star with a photometric and spectroscopic period of $\sim$37 days.  They were able to produce the period but the model light amplitude was too large ($\sim$0.4 mag compared to the observed 0.06$-$0.15 mag) and the fit required too low a temperature (5500 K rather 7250 K) and too high a mass (0.8 M$_{\sun}$ rather than the expected 0.6 M$_{\sun}$.)
The same problems would exist in applying their model to IRAS 17436+5003 and 19475+3119, and for IRAS 18095+2704 the period is longer than their models can account for.
\citet{aik10} computed radial models for post-AGB stars of temperatures in the range of 5000 to 7100 K.  His model for {\it T$_{eff}$} = 7100 K (log {\it T$_{eff}$} = 3.85) was stable in the fundamental and the first several overtones and produced a fundamental period of 32 days (at log {\it g} = 0.5).  He ran nonlinear models at various temperatures.  However, the results for the best fit in temperature (log {\it T$_{eff}$} = 3.85) produced far too short a period ($<$ 10 days) and much too small an amplitude ($<$ 0.01 mag).  Models at lower temperatures (log {\it T$_{eff}$} = 3.75 or {\it T$_{eff}$} = 5600 K) produced more suitable amplitudes ($\sim$0.2 peak to peak), but the periods were still too short ($<$ 20 days).

\section{SUMMARY AND CONCLUSIONS}

In this study, we carried out radial velocity monitoring of three PPNe , IRAS 17436+5003, 18095+2704, and 19475+3119, over intervals of five (1991$-$1995) and nine (2007$-$2015) years, along with photometric BVR$_C$ monitoring over the later interval.
The targets were all of F spectral types.

1. For each of the three, the light and velocity curves had similar periods, which ranged from 35 to 103 days.

2. The pulsation amplitudes are not large, with peak-to-peak ranges of variation of $\le$0.22 mag in {\it V}, $\le$0.10 mag in ({\it B$-$V}), and $\le$16 km~s$^{-1}$ in {\it V$_R$}.

3. For all three objects plus two previously-studies PPNe, (a) the light and color curves are approximately in phase, with the suggestion that the color curve perhaps peaks slightly ($\sim$0.05 {\it P}) after the light. 
(b) The radial velocity curve is approximately $-$0.25 {\it P} out of phase with the light curve.

4. The pulsation characteristics are such that the star is brightest when smallest and hottest, with the temperature maximum occurring slightly after minimum size as the star starts to expand.

With the results of this and our previous study \citep{hri13}, good observational pulsation properties $-$ period, amplitudes of the light and velocity curves, phase difference between the light and velocity $-$ exist for five 
of the brightest PPNe.
Pulsational models can be compared to these observational results for different values of effective temperature, mass, and luminosity.  The temperatures of each of these five PPNe have been determined observationally from high-resolution spectroscopic studies.  With the upcoming release of parallax measurements from the Gaia mission, we will for the first time be able to determine direct distances for these objects and thus their luminosities.  Thus pulsational models have the potential to provide the first opportunity to determine masses of these PPNe (none are in know binary systems  \citep{hri17} and even then, they would almost certainly be single-line) and test post-AGB evolutionary models.
The important need for the construction of new pulsational models is obvious, and we close with an appeal to the stellar pulsation modeling community to address this challenging but rewarding situation.

\section{APPENDIX I. DETERMINATION OF THE EMPIRICAL RADIAL VELOCITY OFFSETS}

As mentioned earlier, when compared carefully, the contemporaneous DAO-CCD, HERMES, and CORAVEL radial velocity sets possessed systematic differences.  This was readily apparent when plotting the radial velocities as measured.  We determined the value of the systematic difference empirically for each star using one of several means, depending upon which one gave the most reliable results.
The radial velocity sets were then adjusted by these differences (offsets), which were calculated in reference to the HERMES values.
The first and most reliable method was to do a period analysis of the individual data sets for a star and compare the systemic velocities.
For IRAS 18095+2704, an analysis of the 2007$-$2015 DAO-CCD velocities and an analysis of the 2009-2015 HERMES radial velocities each led to a period 103.5 days, although each period was slightly below the significance criteria.   
The systemic velocities derived from the periodic fit to each of these two data sets are listed in Table~\ref{rv_results}.
Comparing the systemic velocity of each data set led to an offset of the DAO-CCD radial velocities with respect to the HERMES velocities of $-$1.45 km~s$^{-1}$.
Only for IRAS 18095+2704 did the different data sets possess similar period values, and thus we needed to resort to other means to determine the offsets for the other two stars.
For IRAS 19475+3119, we determined an offset of $-$1.2 km~s$^{-1}$.
This was based on the difference between the systemic velocity of the periodic fit to the HERMES data to the average velocity of the DAO-CCD data (for which no periodic fit was found).
These two values can also be found in Table~\ref{rv_results}.
For IRAS 17436+5003, the object with the greatest number of observations, the values of the systematic offsets were reliably determined based on observations of the objects on the same or on adjacent nights with the different telescope systems.  This resulted in offsets (HERMES minus system) of $-$0.7 km~s$^{-1}$ for DAO-CCD and $+$0.5 km~s$^{-1}$ for CORAVEL, based on 16 and 11 pairs of radial velocities, respectively.  

\acknowledgments 
We gratefully acknowledge the work of D. Bohlender in organizing and reducing the DAO-CCD radial velocity data. We thank R. Manick for providing information on the phasing of two RV Tauri variables, T. Hillwig for a helpful conversation, and the referee for a careful reading of the manuscript and helpful suggestions especially as regards the presentation.
We also thank the many Valparaiso University undergraduate students who carried out the photometric observations used in this study, most recently 
Allyse Appel, Brendan Ferris, Justin Reed, Jacob Bowman, Ryan Braun, Dani Crispo, Stephen Freund,  Chris Morrissey, Cole Hancock, and Abigail Vance, who observed from 2013 to 2015.
BJH acknowledges ongoing support from the National Science Foundation
(most recently AST 1009974, 1413660).  
JS acknowledge support of the Research Council of Lithuania under the grant MIP-085/2012.
This research is based in part on observations obtained at the Dominion Astrophysical Observatory, NRC Herzberg, Programs in Astronomy and Astrophysics, National Research Council of Canada.
It is also based in part on observations made using the Mercator Telescope, operated on the island of La Palma by
the Flemish Community, situated at the Spanish Observatorio del Roque delos Muchachos of the Instituto de Astro\'{f}isica de Canarias. We used data from the HERMES spectrograph, supported by the Fund for Scientific Research of Flanders (FWO), the Research Council of K.U. Leuven, the Fonds National Recherches Scientific (FNRS), the Royal Observatory of Belgium, the Observatoire de Gen\`{e}ve, Switzerland, and the Thringer Landessternwarte Tautenburg, Germany.  This research has been conducted in part based on funding from the Research Council of K.U. Leuven (GOA/13/012) and was partially funded by the Belgian Science Policy Office under contract BR/143/A2/STARLAB.
This research has made use of the SIMBAD database, operated at CDS, Strasbourg,
France, and NASA's Astrophysical Data System.

\clearpage

\tablenum{1}
\begin{deluxetable}{crcclrrrccl}
\tablecaption{Program Objects\label{object_list}}
\tabletypesize{\footnotesize} 
\tablewidth{0pt} \tablehead{
\colhead{IRAS ID}&\colhead{V\tablenotemark{a}}&\colhead{(B$-$V)\tablenotemark{a,b}}
&\colhead{(V$-$R)\tablenotemark{a,b}}&\colhead{Sp.T.}&\colhead{T$_{\rm eff}$}
&\colhead{log {\it g}}&\colhead{[Fe/H]}
&\colhead{[C/O]}&\colhead{Ref.\tablenotemark{c}}&\colhead{Other ID}\\
\colhead{}&\colhead{(mag)}&\colhead{(mag)}&\colhead{(mag)}&\colhead{}
&\colhead{(K)}&\colhead{}&\colhead{}&\colhead{}&\colhead{}&\colhead{}} \startdata
17436+5003 & 7.0 & 0.9 & 0.2 & F3 Ib & 6600 & 0.0 & $-$0.3 & $-$0.2 &1 & HD 161796, V814 Her \\
                   & &  & &  & 7100 & 0.5 & $-$0.2 & $-$0.3 &2 & \nodata \\
                   & &  & &  & 7200 & 1.1 & $-$0.3 & $-$0.2 &3 & \nodata \\
18095+2704 & 10.0 & 1.0 & 0.7 & F3 Ib & 6500 & 0.5 & $-$0.9 & $-$0.4 &4 & HD 335675, V887 Her \\
19475+3119 &  9.3 & 0.6 & 0.4 & F3 Ib & 7750 & 1.0 & $-$0.2 & $-$0.4 &5 & HD 331319, V2513 Cyg  \\
                   & & & & & 7200 & 0.5 & $-$0.2 & $-$0.5 &2\tablenotemark{d} & \nodata \\
\enddata
\tablenotetext{a}{Variable. }
\tablenotetext{b}{Includes circumstellar and interstellar reddening. }
\tablenotetext{c}{References for the spectroscopic analyses: (1) \citet{luck90}, (2) \citet{kloch02}, (3) \citet{tak07}, (4) \citet{sah11}, (5) \citet{ferro01}.}
\tablenotetext{d}{Several He lines measured, suggesting He overabundance \citep{kloch02}. }
\end{deluxetable}


\tablenum{2}
\begin{deluxetable}{clrrcl}
\tablecaption{Summary of the Radial Velocity Data Sets\label{data_sets}}
\tabletypesize{\footnotesize} 
\tablewidth{0pt} \tablehead{
\colhead{IRAS ID}&\colhead{Data Set}
&\colhead{Years} & \colhead{No.}&\colhead{$<$Vo$>$} &\colhead{$\Delta$V$_R$(offsets)\tablenotemark{a}}\\
\colhead{}&\colhead{}&\colhead{}&\colhead{Obs.}
&\colhead{(km~sec$^{-1}$)}&\colhead{(km~sec$^{-1}$)} } 
\startdata
17436+5003 & Burki\tablenotemark{b}         & 1978$-$1979 & 74 & $-$52.34 & \nodata \\
                   & DAO-RVS & 1991$-$1995 & 59 & $-$53.06 & \nodata \\
                   & DAO-CCD & 2007$-$2015 &121 & $-$51.99 & $-$0.7 \\
                   & CORAVEL & 2008$-$2014 & 105 & $-$53.27 & $+$0.5 \\
                   & HERMES       & 2009$-$2015 & 107 & $-$53.07 & \nodata \\
18095+2704  & DAO-RVS & 1991$-$1995 & 47 & $-$29.36 & \nodata \\
                   & DAO-CCD & 2007$-$2015 & 78 & $-$30.32 & $-$1.45\\
                   & HERMES       & 2009$-$2015 & 75 & $-$31.58 & \nodata \\
19475+3119  & DAO-RVS & 1991$-$1995 & 38 & 2.13 & \nodata \\
                   & DAO-CCD & 2007$-$2015 & 77 & 2.18 & $-$1.2 \\
                   & HERMES       & 2009$-$2015 & 57 & 0.94 &  \nodata \\
\enddata
\tablecomments{The precision in the HERMES velocities is $\pm$0.3$-$0.5 km~s$^{-1}$ and the precision in the other velocities is $\pm$0.5$-$0.7 km~s$^{-1}$. }
\tablenotetext{a}{Radial velocity offsets for individual objects needed to bring the DAO-CCD and CORAVEL velocities into agreement with the HERMES velocities: $\Delta$V$_R$ = V$_R$(HERMES) $-$ V$_R$(DAO-CCD) or V$_R$(HERMES) $-$ V$_R$(CORAVEL). 
For DAO-RVS and \citet{bur80}, these empirical offsets are unknown since the observations were not made concurrently.}
\tablenotetext{b}{\citet{bur80}.}
\end{deluxetable}

\clearpage

\tablenum{3}
\begin{deluxetable}{cccccccc}
\tablecaption{Radial Velocity Observations of
IRAS17436+5003\tablenotemark{a}
\label{rvdata_17436}}
\tabletypesize{\scriptsize}
\tablewidth{0pt} \tablehead{ \colhead{HJD$-$2,400,000}
&\colhead{V$_{\rm r}$} 
&&\colhead{HJD$-$2,400,000} &\colhead{V$_{\rm r}$} 
&&\colhead{HJD$-$2,400,000}
&\colhead{V$_{\rm r}$} \\
\colhead{}&\colhead{(km s$^{-1}$)}&&\colhead{}&\colhead{(km s$^{-1}$)}&&\colhead{}&\colhead{(km s$^{-1}$)}  }
\startdata
&&& DAO-RVS &&&& \\
\tableline
48431.8452 & -56.05 && 48733.8158 & -54.50 && 49217.8365 & -55.85 \\
48432.7761 & -55.28 && 48736.8185 & -54.23 && 49239.7111 & -51.52 \\
48450.7673 & -51.90 && 48755.8660 & -51.54 && 49243.7973 & -52.34 \\
48451.7955 & -51.48 && 48756.8473 & -51.60 && 49260.7975 & -53.69 \\
48469.8170 & -55.19 && 48769.8769 & -52.82 && 49263.7786 & -53.85 \\
48471.8429 & -56.21 && 48778.8357 & -54.29 && 49286.7754 & -52.69 \\
48481.7421 & -55.31 && 48778.8417 & -54.46 && 49385.0350 & -55.10 \\
48482.7223 & -55.38 && 48791.7768 & -52.91 && 49443.9042 & -51.62 \\
48510.7929 & -54.35 && 48792.7660 & -53.08 && 49483.9005 & -49.66 \\
48530.6703 & -52.23 && 48799.7755 & -51.21 && 49492.8657 & -54.25 \\
48531.6603 & -53.14 && 48834.7442 & -52.34 && 49525.7946 & -48.14 \\
48532.7120 & -53.97 && 48836.7691 & -52.23 && 49554.7451 & -50.90 \\
48533.7392 & -54.45 && 48875.6629 & -53.34 && 49555.7680 & -51.23 \\
48547.7133 & -52.22 && 48886.7415 & -53.40 && 49800.9470 & -53.18 \\
48549.6874 & -52.26 && 49129.9427 & -50.74 && 49801.9149 & -51.58 \\
48551.7548 & -53.06 && 49130.8159 & -51.14 && 49802.9878 & -52.38 \\
48566.6907 & -53.96 && 49146.9091 & -52.14 && 49913.8331 & -53.98 \\
48588.6521 & -54.14 && 49161.7826 & -53.35 && 49941.8629 & -51.06 \\
48714.9183 & -52.31 && 49163.8331 & -53.62 && 50023.6486 & -53.45 \\
48719.8251 & -55.61 && 49177.8616 & -52.43 && \nodata & \nodata \\
\tableline
&&& DAO-CCD &&&& \\
\tableline
54286.7575 & -51.95 && 55432.7577 & -47.31 && 56548.6539 & -51.12 \\
54297.7496 & -51.64 && 55476.7054 & -51.07 && 56562.6297 & -54.16 \\
54308.8085 & -54.45 && 55482.7469 & -50.42 && 56579.6051 & -50.04 \\
54327.6931 & -52.07 && 55524.6579 & -49.61 && 56582.5995 & -48.36 \\
54338.6991 & -50.50 && 55670.8507 & -48.74 && 56664.0941 & -51.04 \\
54350.7633 & -54.81 && 55691.8283 & -55.42 && 56772.7824 & -48.18 \\
54354.6516 & -56.83 && 55719.7481 & -50.83 && 56778.7858 & -49.19 \\
54384.6009 & -49.62 && 55727.7386 & -49.21 && 56790.7581 & -54.73 \\
54397.5873 & -53.77 && 55812.6711 & -51.69 && 56811.7349 & -53.03 \\
54403.6143 & -55.55 && 55866.7226 & -50.26 && 56829.7413 & -53.81 \\
54434.6422 & -51.08 && 55881.6239 & -51.93 && 56831.7413 & -53.44 \\
54586.9266 & -56.30 && 55908.5869 & -53.07 && 56854.7401 & -55.60 \\
54614.8325 & -54.04 && 55916.5715 & -49.66 && 56876.7135 & -50.27 \\
54626.8112 & -51.45 && 55973.0704 & -54.34 && 56881.7063 & -53.17 \\
54658.7473 & -52.37 && 55981.0255 & -51.70 && 56887.6941 & -53.13 \\
54669.7454 & -50.96 && 56022.8821 & -52.86 && 56904.6670 & -51.59 \\
54690.7359 & -53.68 && 56040.9423 & -46.21 && 56908.6637 & -51.60 \\
54709.7012 & -52.09 && 56076.8267 & -49.65 && 56910.6580 & -51.01 \\
54719.6815 & -52.66 && 56133.7712 & -51.93 && 56913.6527 & -50.92 \\
54725.6725 & -52.59 && 56135.7819 & -50.42 && 56969.6338 & -50.90 \\
54754.6181 & -52.97 && 56142.7727 & -47.62 && 56971.6425 & -53.31 \\
54925.8223 & -50.17 && 56144.7550 & -46.56 && 56972.6301 & -52.60 \\
54944.8169 & -54.84 && 56145.7643 & -46.43 && 56973.6288 & -53.62 \\
54945.8219 & -55.03 && 56160.7584 & -52.02 && 57052.9960 & -51.74 \\
55014.7532 & -54.90 && 56173.7507 & -56.09 && 57142.8297 & -51.42 \\
55028.7185 & -56.48 && 56174.7665 & -55.67 && 57150.7872 & -53.52 \\
55041.7179 & -53.18 && 56185.7528 & -51.07 && 57154.7684 & -53.24 \\
55064.7044 & -53.63 && 56187.7455 & -51.27 && 57183.7457 & -51.95 \\
55076.7077 & -52.13 && 56206.7185 & -54.62 && 57227.7316 & -53.42 \\
55126.6499 & -54.07 && 56240.6494 & -50.69 && 57234.7257 & -51.57 \\
55167.5804 & -51.24 && 56246.6790 & -50.77 && 57239.7184 & -50.58 \\
55288.0344 & -50.83 && 56349.9882 & -54.49 && 57241.7119 & -51.21 \\
55301.7999 & -55.44 && 56370.9204 & -51.98 && 57243.7098 & -50.71 \\
55309.7885 & -55.55 && 56398.8321 & -52.40 && 57245.7108 & -51.86 \\
55322.7482 & -52.41 && 56404.8340 & -52.71 && 57247.7072 & -51.33 \\
55337.7256 & -49.01 && 56413.7722 & -52.53 && 57257.6905 & -53.69 \\
55364.7431 & -54.96 && 56432.7358 & -51.94 && 57316.6413 & -52.28 \\
55385.7409 & -46.74 && 56487.7603 & -51.02 && 57336.6197 & -50.52 \\
55392.7375 & -47.24 && 56498.7430 & -50.30 && 57355.6001 & -53.54 \\
55406.7234 & -58.11 && 56530.7122 & -48.78 && \nodata & \nodata \\
55426.7731 & -49.16 && 56543.6581 & -51.54 && \nodata & \nodata \\
\tableline
&&& HERMES &&&& \\
\tableline
54934.7271 & -55.13 && 56082.5809 & -55.80 && 56695.7552 & -51.31 \\
55022.4583 & -57.63 && 56088.4760 & -54.30 && 56696.7588 & -51.48 \\
55044.4354 & -53.88 && 56111.5765 & -55.99 && 56697.7882 & -51.54 \\
55064.3872 & -54.80 && 56117.3834 & -54.95 && 56698.7672 & -51.62 \\
55231.7647 & -53.93 && 56136.5554 & -49.83 && 56699.7421 & -51.67 \\
55298.5576 & -55.37 && 56148.4501 & -48.16 && 56700.7681 & -51.50 \\
55322.7337 & -52.80 && 56158.3793 & -53.64 && 56701.7266 & -51.00 \\
55353.4440 & -53.77 && 56174.4012 & -57.08 && 56755.6427 & -55.71 \\
55374.4919 & -55.23 && 56174.4051 & -57.06 && 56762.6620 & -51.03 \\
55427.3865 & -49.69 && 56179.4342 & -54.32 && 56767.5336 & -48.33 \\
55444.3820 & -50.94 && 56192.4690 & -50.22 && 56780.5187 & -51.25 \\
55614.6741 & -53.61 && 56307.7705 & -58.87 && 56797.5724 & -52.60 \\
55623.7365 & -51.45 && 56316.7397 & -55.30 && 56809.4447 & -54.61 \\
55638.6396 & -52.95 && 56326.7771 & -52.98 && 56818.4279 & -54.62 \\
55646.7511 & -58.76 && 56334.7507 & -53.15 && 56822.4419 & -55.09 \\
55650.6600 & -54.74 && 56348.7008 & -54.34 && 56838.4963 & -52.95 \\
55652.7219 & -53.29 && 56374.7227 & -54.05 && 56841.5245 & -53.55 \\
55664.6956 & -49.79 && 56381.5880 & -56.60 && 56842.5038 & -53.63 \\
55678.5331 & -52.23 && 56395.6373 & -52.83 && 56843.4884 & -53.65 \\
55684.7586 & -58.89 && 56417.5598 & -54.12 && 56844.5172 & -53.58 \\
55689.5130 & -55.49 && 56424.4764 & -54.35 && 56878.4931 & -53.54 \\
55700.5623 & -52.20 && 56439.5115 & -51.65 && 56882.4087 & -55.35 \\
55710.4695 & -53.19 && 56440.5925 & -51.51 && 56971.3061 & -53.26 \\
55718.5874 & -51.35 && 56446.5949 & -48.46 && 57049.8016 & -53.41 \\
55758.6305 & -53.52 && 56451.4959 & -48.62 && 57098.6353 & -53.86 \\
55829.3542 & -54.90 && 56452.4737 & -48.95 && 57108.7359 & -55.71 \\
55942.7576 & -53.62 && 56454.5546 & -49.33 && 57135.7484 & -47.69 \\
55959.7107 & -52.25 && 56455.5342 & -49.39 && 57151.5634 & -54.02 \\
55966.7644 & -53.26 && 56487.3878 & -51.94 && 57154.5044 & -53.80 \\
55993.6352 & -49.61 && 56525.3944 & -51.31 && 57154.6584 & -53.69 \\
56004.7714 & -57.10 && 56544.4202 & -51.15 && 57191.5097 & -54.28 \\
56009.7334 & -56.42 && 56614.2934 & -51.00 && 57212.4333 & -53.92 \\
56031.5251 & -48.88 && 56679.7679 & -52.18 && 57231.4039 & -53.15 \\
56041.6474 & -48.84 && 56692.7230 & -51.30 && 57251.4260 & -51.92 \\
56045.6703 & -56.86 && 56693.7645 & -51.39 && 57276.4364 & -52.39 \\
56057.4836 & -56.98 && 56694.7185 & -51.39 && \nodata & \nodata \\
\tableline
&&& CORAVEL &&&& \\
\tableline
54747.234 & -52.80 && 55267.656 & -57.00 && 56035.598 & -47.40 \\
54748.250 & -53.10 && 55271.648 & -55.70 && 56091.500 & -53.50 \\
54749.203 & -53.50 && 55301.570 & -55.30 && 56098.387 & -52.60 \\
54764.211 & -52.40 && 55302.594 & -56.40 && 56107.516 & -52.20 \\
54814.180 & -54.10 && 55346.504 & -48.90 && 56315.707 & -54.80 \\
54815.137 & -54.40 && 55365.465 & -56.60 && 56349.664 & -54.40 \\
54864.676 & -51.50 && 55394.484 & -50.20 && 56359.531 & -53.80 \\
54865.676 & -50.90 && 55445.254 & -52.90 && 56359.543 & -53.60 \\
54916.621 & -50.10 && 55671.590 & -50.70 && 56379.590 & -54.90 \\
54929.516 & -54.40 && 55673.594 & -50.50 && 56391.582 & -54.10 \\
54930.488 & -54.20 && 55675.566 & -50.60 && 56398.598 & -53.20 \\
54940.523 & -54.80 && 55693.539 & -57.50 && 56399.594 & -52.90 \\
54941.520 & -56.60 && 55696.531 & -55.00 && 56404.590 & -53.20 \\
54944.488 & -55.60 && 55707.535 & -52.50 && 56542.430 & -51.60 \\
54947.527 & -55.30 && 55708.523 & -53.30 && 56629.207 & -47.80 \\
54950.543 & -53.40 && 55719.516 & -51.30 && 56680.730 & -53.30 \\
54953.547 & -53.80 && 55755.418 & -54.00 && 56681.730 & -52.60 \\
54968.406 & -50.20 && 55801.438 & -53.20 && 56683.723 & -53.50 \\
55000.402 & -50.10 && 55808.344 & -51.60 && 56693.707 & -52.20 \\
55013.394 & -52.60 && 55815.305 & -51.50 && 56713.656 & -56.50 \\
55027.363 & -57.40 && 55816.340 & -53.00 && 56727.633 & -56.00 \\
55030.379 & -55.70 && 55826.277 & -58.20 && 56752.625 & -60.70 \\
55037.391 & -51.60 && 55836.230 & -54.30 && 56758.606 & -56.30 \\
55069.359 & -53.90 && 55853.191 & -51.40 && 56764.602 & -51.60 \\
55076.391 & -50.50 && 55860.223 & -51.50 && 56770.586 & -48.40 \\
55077.324 & -51.20 && 55881.262 & -52.30 && 56773.574 & -51.80 \\
55092.305 & -52.10 && 55968.703 & -54.60 && 56776.570 & -52.10 \\
55093.336 & -52.10 && 55969.711 & -54.50 && 56792.551 & -55.40 \\
55095.328 & -51.40 && 55974.707 & -53.60 && 56814.512 & -52.90 \\
55120.293 & -54.70 && 55994.664 & -49.20 && 56819.426 & -55.60 \\
55124.176 & -54.60 && 55995.648 & -50.10 && 56821.383 & -54.60 \\
55138.215 & -52.50 && 56000.656 & -56.90 && 56905.336 & -53.40 \\
55219.688 & -52.60 && 56010.652 & -57.60 && 56916.328 & -53.20 \\
55220.699 & -52.50 && 56023.621 & -53.10 && 56932.254 & -51.80 \\
55266.641 & -55.90 && 56029.606 & -49.70 && 56954.234 & -54.10 \\
\enddata
\tablenotetext{a}{These are the radial velocities as measured on the individual spectrograph-detector systems, without the inclusion of the empirically-determined offsets.  See the text for details and offset values.} 
\end{deluxetable}

\clearpage

\tablenum{4}
\begin{deluxetable}{cccccccc}
\tablecaption{Radial Velocity Observations of
IRAS 18095+2704\tablenotemark{a}
\label{rvdata_18095}}
\tabletypesize{\scriptsize}
\tablewidth{0pt} \tablehead{ \colhead{HJD$-$2,400,000}
&\colhead{V$_{\rm r}$} 
&&\colhead{HJD$-$2,400,000} &\colhead{V$_{\rm r}$} 
&&\colhead{HJD$-$2,400,000}
&\colhead{V$_{\rm r}$} \\
\colhead{}&\colhead{(km s$^{-1}$)}&&\colhead{}&\colhead{(km s$^{-1}$)}&&\colhead{}&\colhead{(km s$^{-1}$)}  }
\startdata
&&& DAO-RVS &&&& \\
\tableline
48469.7901 & -27.47 && 48733.8520 & -33.02 && 49217.7120 & -30.90 \\
48469.9219 & -27.23 && 48736.8545 & -33.60 && 49239.7479 & -27.19 \\
48470.7376 & -27.70 && 48755.8858 & -31.48 && 49243.8352 & -25.48 \\
48470.8601 & -27.34 && 48756.8639 & -31.02 && 49263.7443 & -28.05 \\
48470.9171 & -27.35 && 48769.9043 & -30.44 && 49265.6606 & -28.62 \\
48471.7762 & -27.31 && 48778.8924 & -30.01 && 49286.6543 & -32.33 \\
48471.8321 & -28.27 && 48788.8935 & -27.62 && 49492.9231 & -30.43 \\
48510.7795 & -30.01 && 48791.8042 & -27.33 && 49525.8683 & -26.95 \\
48511.7727 & -30.03 && 48792.7864 & -28.50 && 49554.7667 & -29.10 \\
48532.7000 & -32.92 && 48799.8455 & -29.16 && 49556.8112 & -28.87 \\
48533.7271 & -33.74 && 48834.7540 & -28.28 && 49801.0325 & -28.91 \\
48551.6795 & -28.47 && 48875.6836 & -33.95 && 49913.8542 & -30.07 \\
48566.6581 & -29.28 && 49130.8524 & -26.64 && 49940.8925 & -31.47 \\
48588.6009 & -26.43 && 49146.8838 & -25.98 && 49941.8904 & -32.27 \\
48714.9559 & -30.11 && 49168.8834 & -28.34 && 50023.6226 & -28.11 \\
48719.9151 & -30.79 && 49189.7655 & -31.12 && \nodata & \nodata \\
\tableline
&&& DAO-CCD &&&& \\
\tableline
54327.7247 & -29.34 && 55392.7803 & -31.58 && 56562.7149 & -27.32 \\
54354.6831 & -29.98 && 55406.7671 & -29.82 && 56579.6898 & -30.40 \\
54391.6727 & -29.75 && 55426.8046 & -25.80 && 56582.6845 & -28.95 \\
54403.6657 & -30.65 && 55719.7942 & -29.71 && 56811.7848 & -31.22 \\
54658.7846 & -30.69 && 55812.7130 & -32.59 && 56829.7867 & -28.33 \\
54669.7755 & -35.48 && 55881.5849 & -31.39 && 56831.7868 & -29.43 \\
54709.7316 & -31.19 && 56133.8041 & -29.37 && 56854.7902 & -29.96 \\
54719.7116 & -32.38 && 56135.8149 & -29.09 && 56881.7560 & -27.94 \\
54725.7023 & -32.22 && 56142.8055 & -27.92 && 56887.7443 & -28.30 \\
54790.5789 & -28.93 && 56144.8187 & -28.68 && 56904.7500 & -31.72 \\
54925.9244 & -30.01 && 56160.8263 & -29.58 && 56908.7466 & -30.25 \\
54945.8753 & -31.96 && 56173.8183 & -29.23 && 56910.7408 & -30.02 \\
55014.7961 & -29.71 && 56174.8340 & -27.05 && 56913.7354 & -31.58 \\
55028.7614 & -29.14 && 56185.7863 & -29.60 && 56915.7819 & -31.01 \\
55041.7607 & -31.61 && 56187.7788 & -31.46 && 57053.0422 & -30.70 \\
55050.7476 & -30.42 && 56240.6156 & -29.70 && 57150.8364 & -31.83 \\
55055.7282 & -31.02 && 56350.0348 & -33.00 && 57183.7957 & -27.73 \\
55063.7145 & -31.08 && 56398.8786 & -30.35 && 57227.7818 & -30.55 \\
55076.7577 & -30.16 && 56404.8807 & -33.85 && 57234.7757 & -30.85 \\
55096.6587 & -30.07 && 56413.8208 & -33.70 && 57239.7683 & -29.66 \\
55167.6077 & -30.77 && 56432.7885 & -30.45 && 57241.7618 & -31.60 \\
55301.8657 & -30.13 && 56487.8101 & -30.26 && 57245.7606 & -30.71 \\
55309.8548 & -29.21 && 56498.7927 & -32.85 && 57257.7740 & -30.60 \\
55322.7971 & -30.43 && 56530.7615 & -29.92 && 57316.6941 & -31.46 \\
55337.7794 & -29.22 && 56543.7070 & -28.21 && 57336.6719 & -31.19 \\
55385.7840 & -31.77 && 56548.7401 & -28.43 && 57355.6366 & -30.64 \\
 \tableline
&&& HERMES &&&& \\
\tableline
55023.6225 & -31.17 && 56054.5244 & -31.09 && 56758.6870 & -30.64 \\
55046.4953 & -33.53 && 56069.4778 & -34.55 && 56772.5903 & -31.94 \\
55083.3917 & -32.91 && 56102.4496 & -32.62 && 56780.5382 & -34.62 \\
55106.4217 & -31.67 && 56116.6172 & -31.16 && 56810.4673 & -31.27 \\
55415.4969 & -29.58 && 56129.6579 & -30.01 && 56837.4787 & -32.10 \\
55420.4348 & -28.81 && 56141.4931 & -30.34 && 56839.4556 & -32.12 \\
55426.3998 & -28.25 && 56155.4335 & -30.94 && 56845.4777 & -32.51 \\
55433.3882 & -29.55 && 56179.4418 & -33.19 && 56868.4922 & -31.15 \\
55617.7105 & -32.38 && 56189.3675 & -34.03 && 56878.5732 & -31.54 \\
55689.5503 & -31.41 && 56335.7639 & -31.00 && 56882.4195 & -31.42 \\
55702.4478 & -31.17 && 56380.6472 & -30.32 && 56960.3249 & -30.69 \\
55711.4648 & -30.62 && 56400.5856 & -32.45 && 56971.3390 & -31.69 \\
55745.5527 & -30.56 && 56418.5071 & -34.71 && 57099.6493 & -32.45 \\
55758.6533 & -33.38 && 56439.5616 & -31.73 && 57115.7594 & -31.70 \\
55777.5900 & -32.72 && 56446.7286 & -30.54 && 57130.7420 & -31.27 \\
55778.5000 & -32.98 && 56451.5439 & -30.14 && 57139.6884 & -31.53 \\
55795.3737 & -33.21 && 56452.5317 & -30.04 && 57150.6144 & -32.74 \\
55801.4237 & -33.48 && 56453.5561 & -30.35 && 57165.6708 & -30.04 \\
55829.4348 & -30.41 && 56454.6006 & -30.13 && 57191.5366 & -30.82 \\
55959.7872 & -31.74 && 56455.5825 & -30.31 && 57205.5039 & -33.31 \\
55964.7859 & -33.38 && 56456.5224 & -30.40 && 57228.3854 & -33.21 \\
55993.6860 & -32.74 && 56692.7749 & -30.52 && 57246.4955 & -33.02 \\
56008.7193 & -31.70 && 56693.7393 & -30.29 && 57432.7373 & -33.29 \\
56027.7504 & -30.36 && 56696.7744 & -29.91 && 57435.7910 & -33.02 \\
56031.5690 & -30.16 && 56698.7562 & -29.71 && 57492.6543 & -32.19 \\
\enddata
\tablenotetext{a}{These are the radial velocities as measured on the individual spectrograph-detector systems, without the inclusion of the empirically-determined offsets.  See the text for details and offset values.} 
\end{deluxetable}

\clearpage

\tablenum{5}
\begin{deluxetable}{cccccccc}
\tablecaption{Radial Velocity Observations of
IRAS 19475+3119\tablenotemark{a}
\label{rvdata_19475}}
\tabletypesize{\scriptsize}
\tablewidth{0pt} \tablehead{ \colhead{HJD$-$2,400,000}
&\colhead{V$_{\rm r}$} 
&&\colhead{HJD$-$2,400,000} &\colhead{V$_{\rm r}$} 
&&\colhead{HJD$-$2,400,000}
&\colhead{V$_{\rm r}$} \\
\colhead{}&\colhead{(km s$^{-1}$)}&&\colhead{}&\colhead{(km s$^{-1}$)}&&\colhead{}&\colhead{(km s$^{-1}$)}  }
\startdata
&&& DAO-RVS &&&& \\
\tableline
48471.8722 &   4.50 && 48755.9488 &   2.50 && 49243.8701 &   1.12 \\
48510.8125 &   2.49 && 48769.9662 &  -2.39 && 49260.8258 &   2.48 \\
48511.8436 &   2.33 && 48778.9084 &   0.23 && 49263.7898 &   4.44 \\
48532.7230 &   1.75 && 48788.9215 &   1.82 && 49263.8155 &   3.71 \\
48533.7756 &   0.76 && 48791.8756 &   2.25 && 49265.7889 &   2.25 \\
48547.7313 &   0.77 && 48792.8493 &   1.91 && 49286.7424 &  -0.37 \\
48551.6991 &   2.25 && 48799.8793 &  -0.33 && 49554.8323 &   7.84 \\
48566.7034 &   5.08 && 48834.7884 &   1.68 && 49555.8313 &   6.71 \\
48588.6654 &   1.53 && 48875.7701 &   0.82 && 49802.0291 &   5.40 \\
48715.0042 &   4.88 && 49130.8788 &   1.56 && 49913.9049 &  -0.56 \\
48719.9541 &   1.94 && 49163.8834 &  -0.16 && 49941.9650 &   0.68 \\
48733.8877 &  -0.80 && 49217.8752 &   5.61 && 50020.7299 &   3.16 \\
48736.8856 &  -0.25 && 49239.8243 &   1.28 && \nodata & \nodata \\
\tableline
&&& DAO-CCD &&&& \\
\tableline
54292.8793 &   4.65 && 55432.8100 &  -1.70 && 56548.7968 &   1.05 \\
54327.8300 &   2.20 && 55476.6738 &   1.17 && 56579.7468 &  -1.53 \\
54338.7707 &   0.06 && 55524.6341 &   0.81 && 56582.7415 &  -0.56 \\
54354.8336 &   3.18 && 55719.8479 &   4.16 && 56661.5797 &   2.61 \\
54384.7038 &   2.08 && 55812.7700 &   4.35 && 56772.8530 &   3.51 \\
54397.6181 &  -0.21 && 55866.7531 &   4.21 && 56811.8248 &   1.69 \\
54403.7132 &   3.60 && 55908.6154 &  -3.17 && 56881.8421 &   0.04 \\
54434.6857 &   0.57 && 55916.5999 &   3.16 && 56887.8305 &  -0.35 \\
54614.9333 &   6.74 && 56133.8752 &   0.18 && 56904.8033 &   7.63 \\
54658.8224 &  -0.59 && 56135.8859 &   0.23 && 56908.8000 &   5.82 \\
54669.8137 &  -2.47 && 56142.8837 &   1.12 && 56910.7943 &   2.67 \\
54709.7962 &   1.72 && 56144.8673 &   2.37 && 56913.7889 &   0.58 \\
54719.8078 &   3.36 && 56145.8766 &   3.22 && 56915.8355 &  -1.61 \\
54725.7988 &   2.35 && 56160.8825 &   3.94 && 56972.6637 &   5.74 \\
54754.6919 &   1.97 && 56174.8905 &  -2.17 && 56973.6624 &   5.59 \\
54925.9968 &   0.16 && 56185.8430 &   4.51 && 56976.5908 &   6.54 \\
54945.9058 &   2.08 && 56187.8356 &   3.96 && 57150.8776 &   3.51 \\
55028.8113 &   3.11 && 56206.8083 &   6.38 && 57183.8376 &   2.90 \\
55041.8120 &   8.48 && 56350.0762 &   9.00 && 57234.8686 &   4.71 \\
55064.8040 &  -0.37 && 56398.9177 &  -1.71 && 57241.8478 &  -5.48 \\
55096.7053 &  -0.67 && 56404.9198 &   2.68 && 57243.8139 &  -3.53 \\
55167.6551 &   4.62 && 56413.8617 &   3.08 && 57245.8466 &  -6.38 \\
55309.9258 &  -2.42 && 56487.8513 &   3.71 && 57257.8415 &   5.66 \\
55322.8473 &   3.34 && 56498.8361 &   6.80 && 57316.7480 &   1.87 \\
55385.8342 &   4.30 && 56530.8429 &   1.18 && 57336.7255 &   5.43 \\
55392.8303 &   2.96 && 56543.8001 &   3.69 && \nodata & \nodata \\
\tableline
&&& HERMES &&&& \\
\tableline
55013.6044 &   2.44 && 55842.3462 &   1.58 && 56815.5749 &  -3.39 \\
55013.6259 &   2.31 && 55883.3688 &   1.54 && 56821.7083 &   2.92 \\
55024.6212 &  -2.65 && 56016.6858 &  -1.75 && 56839.6865 &  -5.38 \\
55033.6197 &   3.52 && 56035.6678 &   3.10 && 56844.6677 &  -4.76 \\
55033.6412 &   3.55 && 56035.6772 &   3.28 && 56871.5711 &   4.73 \\
55083.4950 &   0.92 && 56115.5100 &   4.32 && 56879.4397 &  -2.08 \\
55106.4374 &   3.21 && 56137.5595 &  -3.33 && 56899.5329 &   4.82 \\
55428.6237 &   0.70 && 56159.5545 &   3.99 && 56903.4320 &   6.51 \\
55448.3599 &   0.52 && 56190.5039 &   2.98 && 56903.4464 &   6.53 \\
55499.3846 &   1.53 && 56438.7232 &   6.05 && 56936.4751 &   3.72 \\
55696.7162 &   1.14 && 56463.7188 &  -2.09 && 56970.3905 &   5.98 \\
55715.5326 &  -0.24 && 56483.5995 &  -0.55 && 57131.7438 &  -2.50 \\
55725.5403 &  -3.25 && 56486.5896 &   3.32 && 57150.7205 &   2.91 \\
55759.4742 &  -2.72 && 56602.3881 &   8.13 && 57166.6632 &  -1.16 \\
55763.6719 &  -1.97 && 56620.3479 &  -5.06 && 57192.6296 &   3.30 \\
55784.6330 &   1.07 && 56620.3696 &  -5.12 && 57230.5451 &   3.99 \\
55807.5640 &   2.14 && 56625.3538 &   0.71 && 57250.6105 &  -3.32 \\
55829.4839 &  -2.62 && 56701.7685 &   5.74 && 57354.3669 & -11.06 \\
55840.4100 &   0.67 && 56781.6794 &   0.86 && 57510.7276 &   3.79 \\
\enddata
\tablenotetext{a}{These are the radial velocities as measured on the individual spectrograph-detector systems, without the inclusion of the empirically-determined offsets.  See the text for details and offset values.} 
\end{deluxetable}

\clearpage

\tablenum{6}
\begin{deluxetable}{crrrcrrrr}
\tablecolumns{13} \tabletypesize{\scriptsize}
\tablecaption{Differential Standard {\it BVR$_C$} Magnitudes for IRAS 17436+5003 from 2013$-$2015 
 \label{lcdata_17436}}  
\tablewidth{0pt} \tablehead{\colhead{HJD $-$ 2,400,000\tablenotemark{a}}
&\colhead{$\Delta$B\tablenotemark{b}}  &\colhead{$\Delta$V\tablenotemark{b}} &\colhead{$\Delta$R$_C$\tablenotemark{b}} & &
\colhead{HJD $-$ 2,400,000\tablenotemark{a}}
&\colhead{$\Delta$B\tablenotemark{b}}  &\colhead{$\Delta$V\tablenotemark{b}} &\colhead{$\Delta$R$_C$}\tablenotemark{b}}
\startdata
56436.6855 & -3.867 & -3.324 & -3.138 && 56856.6221 & -3.809 & -3.369 & -3.140 \\
56437.6535 & -3.724 & -3.328 & -3.108 && 56857.5966 & -3.808 & -3.378 & -3.144 \\
56442.6292 & \nodata & -3.324 & -3.107 && 56859.6341 & -3.805 & -3.356 & -3.143 \\
56446.5963 & -3.804 & -3.374 & -3.144 && 56860.5978 & -3.787 & -3.370 & -3.118 \\
56447.6093 & -3.868 & -3.385 & -3.155 && 56862.5922 & -3.804 & -3.355 & -3.131 \\
56451.5946 & -3.865 & -3.388 & -3.188 && 56867.5904 & -3.832 & -3.375 & -3.136 \\
56471.6107 & -3.834 & -3.370 & -3.131 && 56869.5981 & -3.834 & -3.389 & -3.151 \\
56477.7061 & -3.749 & -3.329 & -3.118 && 56870.6079 & -3.833 & -3.362 & -3.126 \\
56483.6781 & -3.779 & -3.343 & -3.114 && 56873.6769 & -3.864 & -3.392 & -3.140 \\
56484.5971 & -3.781 & -3.341 & -3.112 && 56876.6484 & -3.860 & -3.389 & -3.157 \\
\enddata
\tablenotetext{a}{The time represents the mid-time of the {\it V} observations.  The times for the {\it B} and {\it R$_C$} observations differ from this by approximately the following:   +0.0011 and $-$0.0012 days, respectively. }
\tablenotetext{b}{The average uncertainties in brightness are $\pm$0.005 mag.}
\end{deluxetable}


\tablenum{7}
\begin{deluxetable}{crrrcrrrr}
\tablecolumns{13} \tabletypesize{\scriptsize}
\tablecaption{Differential Standard {\it BVR$_C$} Magnitudes for IRAS 18095+2704  from 2013$-$2015 
 \label{lcdata_18095}}  
\tablewidth{0pt} \tablehead{\colhead{HJD $-$ 2,400,000\tablenotemark{a}}
&\colhead{$\Delta$B\tablenotemark{b}}  &\colhead{$\Delta$V\tablenotemark{b}} &\colhead{$\Delta$R$_C$\tablenotemark{b}} & &
\colhead{HJD $-$ 2,400,000\tablenotemark{a}}
&\colhead{$\Delta$B\tablenotemark{b}}  &\colhead{$\Delta$V\tablenotemark{b}} &\colhead{$\Delta$R$_C$}\tablenotemark{b}}
\startdata
56436.7037 & -1.770 & -2.265 & -2.637 && 6842.6365 & -1.954 & -2.391 & -2.744 \\
56437.6672 & -1.769 & -2.270 & -2.644 && 56846.6498 & \nodata & -2.393 & -2.745 \\
56446.6302 & -1.819 & -2.290 & -2.668 && 56847.6608 & -1.983 & -2.388 & -2.752 \\
56447.6181 & -1.816 & -2.298 & -2.684 && 56848.6436 & -1.944 & -2.397 & -2.740 \\
56451.6188 & -1.850 & -2.309 & -2.668 && 56849.6422 & -1.974 & -2.386 & -2.747 \\
56471.6277 & -1.992 & -2.415 & -2.762 && 56853.6858 & -1.941 & -2.373 & -2.732 \\
56483.6873 & -1.951 & -2.391 & -2.739 && 56855.6156 & -1.922 & -2.373 & -2.740 \\
56484.6044 & -1.958 & -2.375 & \nodata && 56856.6360 & \nodata & -2.373 & -2.747 \\
56485.6033 & -1.958 & -2.379 & -2.739 && 56860.6168 & -1.903 & -2.360 & -2.716 \\
56486.7417 & -1.952 & -2.387 & -2.722 && 56862.6661 & -1.871 & -2.353 & -2.705 \\
\enddata
\tablenotetext{a}{The time represents the mid-time of the {\it V} observations.  The times for the {\it B} and {\it R$_C$} observations differ from this by approximately the following: +0.0021 and $-$0.0017 days, respectively.}
\tablenotetext{b}{The average uncertainties in brightness are $\pm$0.005 mag.}
\end{deluxetable}


\tablenum{8}
\begin{deluxetable}{crrrcrrrr}
\tablecolumns{13} \tabletypesize{\scriptsize}
\tablecaption{Differential Standard {\it BVR$_C$} Magnitudes for IRAS 19475+3119  from 2013$-$2015 
 \label{lcdata_19475}}  
\tablewidth{0pt} \tablehead{\colhead{HJD $-$ 2,400,000\tablenotemark{a}}
&\colhead{$\Delta$B\tablenotemark{b}}  &\colhead{$\Delta$V\tablenotemark{b}} &\colhead{$\Delta$R$_C$\tablenotemark{b}} & &
\colhead{HJD $-$ 2,400,000\tablenotemark{a}}
&\colhead{$\Delta$B\tablenotemark{b}}  &\colhead{$\Delta$V\tablenotemark{b}} &\colhead{$\Delta$R$_C$}\tablenotemark{b}}
\startdata
56436.7259 & -1.222 & -1.712 & -2.022 && 56840.6924 & \nodata & -1.714 & -2.032 \\
56437.7096 & -1.240 & -1.736 & -2.056 && 56842.6662 & -1.245 & -1.721 & -2.034 \\
56446.6762 & -1.252 & -1.721 & -2.021 && 56847.6708 & -1.217 & -1.704 & -2.021 \\
56447.6825 & -1.260 & -1.732 & -2.037 && 56848.7743 & -1.245 & -1.694 & -2.012 \\
56471.6339 & -1.223 & -1.708 & -2.022 && 56849.7261 & -1.151 & -1.687 & -1.999 \\
56483.7283 & -1.197 & -1.689 & -2.021 && 56853.7784 & -1.199 & -1.677 & -1.995 \\
56484.6115 & -1.170 & -1.660 & -1.979 && 56855.7698 & -1.209 & -1.689 & -2.003 \\
56485.6082 & -1.179 & -1.674 & -1.986 && 56856.6458 & \nodata & -1.698 & -2.025 \\
56488.5972 & -1.208 & -1.689 & -2.016 && 56857.6449 & -1.210 & -1.694 & -2.012 \\
56489.6135 & -1.223 & -1.704 & \nodata && 56859.6616 & -1.204 & -1.669 & -1.989 \\
\enddata
\tablenotetext{a}{The time represents the mid-time of the {\it V} observations.  The times for the {\it B} and {\it R$_C$} observations differ from this by approximately the following: +0.0027 and $-$0.0018 days, respectively.}
\tablenotetext{b}{The average uncertainties in brightness are $\pm$0.005 mag.}
\end{deluxetable}

\clearpage

\tablenum{9}
\begin{deluxetable}{llrrcrcrrcr} 
\tablecolumns{11} \tabletypesize{\scriptsize}
\tablecaption{Periodogram Study of the Radial Velocity Curves\label{rv_results}}
\rotate
\tabletypesize{\footnotesize} 
\tablewidth{0pt} \tablehead{ 
\colhead{IRAS ID} & \colhead{Data} & \colhead{Years}&\colhead{No.} & \colhead{A$_0$\tablenotemark{a}}& \colhead{P$_1$\tablenotemark{b}}&\colhead{A$_1$} &\colhead{$\phi$$_1$\tablenotemark{c}}&\colhead{P$_2$}& \colhead{A$_2$} &\colhead{$\phi$$_2$\tablenotemark{c}}\\
\colhead{} &\colhead{Set} &\colhead{}&\colhead{Obs.} &\colhead{(km~s$^{-1}$)}& \colhead{(days)}&\colhead{(km~s$^{-1}$)} & &\colhead{(days)}& \colhead{(km~s$^{-1}$)} & }
\startdata
17436+5003 & Burki             & 1978-1979 & 74 & $-$52.44 & 53.4: & 2.06 & 0.89 & \nodata & \nodata & \nodata   \\
17436+5003 & Burki             & 1978 & 17 & $-$51.34 & 50.3 & 3.33 & 0.09 & \nodata & \nodata & \nodata   \\
17436+5003 & Burki             & 1979 & 57 & $-$52.91 & 56.9 & 1.77 & 0.23 & \nodata & \nodata & \nodata   \\
17436+5003 & DAO-RVS     & 1991-1995 & 59 & $-$53.06\tablenotemark{d} & \nodata & \nodata & \nodata & \nodata & \nodata    \\
17436+5003 & DAO-RVS     & 1991-1992 & 34 & $-$53.67 & 42.8 & 1.43 & 0.94 & \nodata & \nodata   \\
17436+5003 & DAO-CCD     & 2007-2015 & 121 & $-$52.26 & 47.7 & 1.81 & 0.76 & \nodata & \nodata & \nodata    \\
17436+5003 & CORAVEL    & 2008-2014 & 105 & $-$53.23 &35.3 & 1.48 & 0.14 & 46.6 & 1.43 & 0.92  \\
17436+5003 & HERMES & 2009-2015 & 107 & $-$53.05 & 51.8: & \nodata & \nodata& \nodata & \nodata & \nodata   \\
17436+5003 & DAO+COR+HER & 2007-2015 & 333 & $-$52.85 & (39.4:) & \nodata & \nodata& \nodata & \nodata & \nodata  \\
17436+5003 & DAO+COR+HER & 2007 & 11 & $-$53.96 & 45.9 & 2.86 & 0.77 & \nodata & \nodata & \nodata    \\
17436+5003 & DAO+COR+HER & 2010 & 32 & $-$52.93 & 50.0 & 3.79 & 0.53 & \nodata & \nodata & \nodata    \\
17436+5003 & DAO+COR+HER & 2011 & 42 & $-$53.21 & 48.6: & 2.16 & 0.83 & \nodata & \nodata & \nodata    \\
17436+5003 & DAO+COR+HER & 2012 & 52 & $-$53.03 & 53.0 & 3.26 & 0.99 & \nodata & \nodata & \nodata    \\
17436+5003 & DAO+COR+HER & 2009-2010 & 70 & $-$52.82 & 47.2: & 2.23 & 0.80 & \nodata & \nodata & \nodata    \\
17436+5003 & DAO+COR+HER & 2012-2013 & 98 & $-$52.78 & (50.3:) &  &  & \nodata & \nodata & \nodata    \\
17436+5003 & DAO+COR+HER & 2012-2013 & 98 & $-$52.72 & 49.52\tablenotemark{e} & 1.80  & 0.32  & \nodata & \nodata & \nodata    \\
\\
18095 & DAO-RVS     & 1991-1995 & 47 & $-$29.68 & 109.2 & 2.31 & 0.46 & \nodata  & \nodata & \nodata  \\
18095 & DAO-CCD     & 2007-2015 & 78 & $-$30.24 & 103.5: & 1.11 & 0.93 & \nodata & \nodata & \nodata    \\
18095 & HERMES          & 2009-2015 & 72 & $-$31.69 & 103.5: & 1.09 & 0.99 & \nodata & \nodata  & \nodata   \\
18095 & DAO-CCD+HER & 2007-2015 & 150 & $-$31.69 & 103.5 & 1.08 & 0.97 & \nodata & \nodata & \nodata   \\
18095 & DAO-CCD+HER & 2010-2013 & 76 & $-$31.74 & 103.4 & 1.61 & 0.95 & \nodata & \nodata & \nodata   \\
18095 & DAO-CCD+HER & 2010-2013 & 76 & $-$31.68 & 102.3\tablenotemark{e} & 1.51 & 0.91 & \nodata & \nodata & \nodata  \\
\\
19475 & DAO-RVS     & 1991$-$1995 & 38 & $+$1.98 & 47.1 & 2.2 & 0.54 & \nodata  & \nodata & \nodata  \\
19475 & DAO-CCD     & 2007$-$2015 & 77 & $+$2.28\tablenotemark{d} & \nodata & \nodata & \nodata & \nodata & \nodata & \nodata    \\
19475 & HERMES          & 2009$-$2015 & 55 & $+$1.07 & 37.2 & 3.0 & 0.38 & \nodata & \nodata & \nodata    \\
19475 & HERMES          & 2009$-$2015 & 55 & $+$0.95 & 38.0 & 3.1 & 0.32 & 33.3 & 2.3 & 0.13    \\
19475  & DAO-CCD+HER    & 2007$-$2015 & 132 & $+$1.26 & 37.1 & 2.1 & 0.31 & \nodata & \nodata & \nodata    \\
19475  & DAO-CCD+HER    & 2012 & 17 & $+$1.45 & 41.6 & 3.3 & 0.90 & \nodata & \nodata & \nodata    \\
19475  & DAO-CCD+HER    & 2014 & 24 & $+$1.47 & 37.1 & 4.2 & 0.19 & \nodata & \nodata & \nodata    \\
19475  & DAO-CCD+HER    & 2015 & 15 & $+$0.13 & 35.3: & 5.0 & 0.14 & \nodata & \nodata & \nodata    \\
19475  & DAO-CCD+HER    & 2009 & 14 & $+$1.68 & 41.84\tablenotemark{e} & 2.2 & 0.64 & \nodata & \nodata & \nodata    \\
19475  & DAO-CCD+HER    & 2015 & 15 & $+$0.15 & 35.04\tablenotemark{e} & 5.1 & 0.83 & \nodata & \nodata & \nodata    \\
\enddata
\tablenotetext{a}{The systemic velocities listed in the table for the individual data sets are as observed, without an offset. 
For combined data sets, offsets are added to the DAO and CORAVEL data sets to bring them to the HERMES system.}
\tablenotetext{b}{Period values that are not quite significant are indicated with a colon; those that are more uncertain are indicated with a colon and also surrounded by parentheses.}
\tablenotetext{c}{The phases are determined based on the epoch of JD 2,455,600.00.}
\tablenotetext{d}{The average velocity is listed in the absence of a systemic velocity determined from a period analysis.}
\tablenotetext{e}{Period fixed at the value determined from the {\it V} light curve during the associated years.}
\end{deluxetable}


\tablenum{10}
\begin{deluxetable}{rcrrrrrrrrr} 
\tablecolumns{11} \tabletypesize{\scriptsize}
\tablecaption{Contemporaneous Light, Color, and Velocity Curve Fits\label{fit_par}}
\rotate
\tabletypesize{\footnotesize} 
\tablewidth{0pt} \tablehead{ 
\colhead{IRAS} & \colhead{Years}& \colhead{P$_1$}&\colhead{A(V)} &\colhead{$\phi$(V)\tablenotemark{a}}&\colhead{A(B$-$V)}& \colhead{$\phi$(B$-$V)\tablenotemark{a}} &\colhead{A(V$-$R$_C$)}&\colhead{$\phi$(V$-$R$_C$)\tablenotemark{a}}& \colhead{A(V$_R$)} &\colhead{$\phi$(V$_R$)\tablenotemark{a}}\\
\colhead{ID} &\colhead{}& \colhead{(days)}&\colhead{(mag)} & &\colhead{(mag)}& &  \colhead{(mag)} & &  \colhead{(km~s$^{-1}$)} & } 
\startdata
17436+5003 & 2012$-$2013 & 49.5 & 0.046 & 0.58 & 0.021 & 0.57        & 0.011 & 0.55         & 1.86 & 0.32 \\
           &             & $\pm$0.2 & $\pm$ 0.004 & $\pm$ 0.01 & $\pm$ 0.003 & $\pm$ 0.02        & $\pm$ 0.002 & $\pm$ 0.03         & $\pm$ 0.36 & $\pm$ 0.03 \\
           & 2010              & 45.6 & 0.052 & 0.18 & 0.019 & 0.15        & 0.009 & 0.19         & 2.88 & 0.98 \\
           &               & $\pm$ 0.5 & $\pm$ 0.005 & $\pm$ 0.02 & $\pm$ 0.003 & $\pm$ 0.03        & $\pm$ 0.003 & $\pm$ 0.04         & $\pm$ 0.51 & $\pm$ 0.02 \\
           & 1991$-$1992\tablenotemark{b} & 43.3 & 0.028 & 0.52 & 0.009 & 0.43        & \nodata & \nodata & 1.39 & 0.14 \\
           &             & $\pm$ 0.1 & $\pm$ 0.002 & $\pm$ 0.01 & $\pm$ 0.001 $\pm$ & 0.02        & \nodata & \nodata         & $\pm$ 0.22 & $\pm$ 0.03 \\
           & 1979              & 59   & 0.043 & 0.68 & \nodata & \nodata & \nodata & \nodata & 1.75 & 0.49 \\
           &      & $\pm$ 2 & $\pm$ 0.004 & $\pm$ 0.02 & \nodata & \nodata        & \nodata & \nodata     & $\pm$ 0.20 & $\pm$ 0.02 \\
18095+2704 & 2010$-$2013 & 102.3 & 0.032 & 0.20 & 0.014 & 0.13          & 0.006 & 0.18         & 1.51 & 0.91 \\
          &      & $\pm$ 0.2 & $\pm$ 0.002 & $\pm$ 0.01 & $\pm$ 0.001 & $\pm$ 0.02        & $\pm$ 0.001 & $\pm$ 0.03         & $\pm$ 0.21 & $\pm$ 0.02 \\
19475+3119 & 2009              & 41.8 & 0.050 & 0.88 & 0.013 & 0.80          & 0.010 & 0.80         & (2.1)\tablenotemark{c} & (0.64)\tablenotemark{c} \\
          &      & $\pm$ 0.6 & $\pm$ 0.006 & $\pm$ 0.02 & $\pm$ 0.002 & $\pm$ 0.03        & $\pm$ 0.002 & $\pm$ 0.03         & $\pm$ 0.88 & $\pm$ 0.05 \\
          & 2015             & 35.0 & 0.041 & 0.18 & \nodata & \nodata & \nodata & \nodata    & 5.10 & 0.83 \\
          &     & $\pm$ 0.4 & $\pm$ 0.005 & $\pm$ 0.02 & \nodata & \nodata & \nodata & \nodata      & $\pm$ 0.71 & $\pm$ 0.02 \\
\enddata
\tablenotetext{a}{The phases are determined based on the epoch of JD 2,455,600.00.}
\tablenotetext{b}{Excluding the later photometric data in 1992, which showed a relatively large increase in brightness and blueness.}
\tablenotetext{c}{The fit of the radial velocity to this period, or to any reasonable period, is poor.}
\end{deluxetable}


\tablenum{11}
\begin{deluxetable}{rcrrrrrrrr}
\tablecaption{Phase Comparisons of the Light, Color, and Radial Velocity Curves\tablenotemark{a}\label{data_phases}}
\tabletypesize{\scriptsize}
\tablewidth{0pt} \tablehead{
\colhead{IRAS ID}&\colhead{Years} & \colhead{P(day)}
&\colhead{$\phi\arcmin$(V)\tablenotemark{b}} &\colhead{$\phi\arcmin$(B$-$V)\tablenotemark{b}} & \colhead{$\phi\arcmin$(V$-$R$_C$)\tablenotemark{b}} &\colhead{$\phi\arcmin$(V$_R$)\tablenotemark{b}}
&\colhead{$\Delta$($\phi\arcmin$(B$-$V))\tablenotemark{c}} & \colhead{$\Delta$($\phi\arcmin$(V$-$R$_C$))\tablenotemark{c}} &\colhead{$\Delta$($\phi\arcmin$(V$_R$))\tablenotemark{c}}}   
\startdata
17436+5003 & 2012$-$2013 & 49.52 & 0.67 & 0.68 & 0.70 & 0.43 & 0.01 & 0.03 & $-$0.24 \\
                     & 2010               &  45.62&  0.07 & 0.10 & 0.06 & 0.77 & 0.03 & $-$0.01 & $-$0.20 \\
                     & 1991$-$1992 &  43.34&  0.73 & 0.82 & \nodata & 0.61 & 0.09 & \nodata & $-$0.12 \\
                     & 1979              &  59.14&  0.57 & \nodata & \nodata & 0.26 & \nodata & \nodata & $-$0.31 \\
18095+2704 & 2010$-$2013 & 102.3 & 0.05 & 0.12 & 0.07 & 0.84 & 0.07 & 0.02 & $-$0.21 \\
19475+3119 & 2009 & 41.84 & 0.37 & 0.45 & 0.45 & 0.11: & 0.08 & 0.08 & $-$0.26: \\
                    & 2015 & 35.04 & 0.07 & \nodata & \nodata & 0.92 & \nodata & \nodata & $-$0.15 \\
\\
22223+4327 & 2009$-$2011 &88.27 & 0.03 & 0.05 & 0.05 & 0.79 & 0.02 & 0.02 & $-$0.24 \\
22272+5435 & 2005$-$2007 &131.9 & 0.74 & \nodata & 0.74 & 0.50 & \nodata & 0.00 & $-$0.24 \\
                     & 2008$-$2009 &131.9 & 0.70 & 0.67 & 0.72 & 0.42 & $-$0.03 & 0.02 & $-$0.28 \\
\enddata
\tablenotetext{a}{The uncertainties in the phase values are listed in Table~\ref{fit_par}.
The uncertainties in the phase differences range from $\pm$0.02$-$0.03, except for $\pm$0.04 for IRAS 19475+3119 $\Delta$($\phi\arcmin$(V$-$R$_C$) in 2009 and $\pm$0.05 for IRAS 17436+5003 $\Delta$($\phi\arcmin$(V$-$R$_C$) in 2010 and for IRAS 19475+3119 $\Delta$($\phi\arcmin$(V$_R$) in 2009.}
\tablenotetext{b}{These are the phases of the minimum light and radial velocity ($\phi\arcmin$) based on sine curve fits to the data.  They were determined from the phases ($\phi$) listed in Table~\ref{fit_par} in the following way. The offset from phase 0.00 was determined by 1.00 $-$ $\phi$, and then this was adjusted by $-$0.25 phase to the time of minimum and, for the light and color curves, by an additional $+$0.50 to account for the inverted y-axis of the magnitude system.}
\tablenotetext{c}{Difference between the phase at minimum ($\phi\arcmin$) of the ({\it B$-$V}), ({\it V$-$R$_C$}) and {\it V$_R$} curves and the {\it V} curve.}
\end{deluxetable}

\clearpage

\clearpage

\clearpage

\clearpage

\clearpage

\clearpage

\clearpage

\end{document}